\documentclass[letterpaper,twocolumn,10pt]{article}
\usepackage{usenix2019_v3}

\usepackage{tikz}
\usepackage{algorithm}
\usepackage{algpseudocode}
\usepackage{xcolor}
\usepackage[skip=5pt]{subcaption}
\usepackage{amsmath,amssymb,amsfonts}
\usepackage{hyperref}
\usepackage{graphicx}
\usepackage{multirow}
\usepackage{amsthm}
\usepackage{balance}
\usepackage{float}
\usepackage{booktabs}
\usepackage{makecell}
\usepackage{colortbl}
\usepackage[T1]{fontenc}
\usepackage{pifont}

\usepackage{enumitem}
\setlist{leftmargin=0.6cm}

\newtheorem{theorem}{Theorem}

\newtheorem{lemma}{Lemma}

\newtheorem{definition}{Definition}
\newtheorem{corollary}{Corollary}
\newtheorem{property}{Property}

\newcommand{\fmgftoc}{-0.2cm}
\newcommand{\fmgctom}{0cm}
\newcommand{\pmgttom}{-0.2cm}
\newcommand{\smgbs}{-0.3cm}

\newcommand{\equsize}{\small}
\newcommand{\equienvs}{\begin{small}}
\newcommand{\equienve}{\end{small}}

\newcommand{\one}{\ding{182}\hspace{0.1cm}}
\newcommand{\two}{\ding{183}\hspace{0.1cm}}

\newcommand{\mi}{\mathrm{MI}}
\newcommand{\notbf}[1]{\noindent\textbf{#1}}

\newcommand{\prob}{\operatorname{Pr}}
\newcommand{\fdp}[1]{\overset{#1-\text{DP}}{\sim}}
\newcommand{\pr}[2]{{\ifx&#1& \prob \else \underset{#1}{\prob} \fi \left[#2\right]}}
\newcommand{\prc}[3]{{\ifx&#1& \prob \else \underset{#1}{\prob } \fi \left( #2 \middle| #3 \right)}}

\newcommand{\sumb}[2]{{\ifx&#1& \prob \else \underset{#1}{\sum} \fi #2}}

\begin{document}
%-------------------------------------------------------------------------------
\thispagestyle{headings}
\markright{\hfill To appear in the 34th Usenix Security Symposium, August 2025, SEATTLE, WA, USA\hfill}

%don't want date printed
\date{}

% make title bold and 14 pt font (Latex default is non-bold, 16 pt)
\title{Privacy Audit as Bits Transmission: (Im)possibilities for Audit by One Run}

%for single author (just remove % characters)
\author{
{\rm Zihang Xiang}\\
{\it KAUST}\\
\textit{zihang.xiang@kaust.edu.sa}
\and
{\rm Tianhao Wang}\\
{\it University of Virginia}\\
\textit{tianhao@virginia.edu}
\and
{\rm Di Wang}\\
{\it KAUST}\\
\textit{di.wang@kaust.edu.sa}
% copy the following lines to add more authors
% \and
% {\rm Name}\\
%Name Institution
} % end author

\maketitle

\maketitle
% \vspace{-1cm}
\begin{abstract}

Auditing algorithms' privacy typically involves simulating a game-based protocol that guesses which of two adjacent datasets was the original input. Traditional approaches require thousands of such simulations, leading to significant computational overhead. Recent methods propose single-run auditing of the target algorithm to address this, substantially reducing computational cost. However, these methods' general applicability and tightness in producing empirical privacy guarantees remain uncertain.

This work studies such problems in detail. Our contributions are twofold: First, we introduce a unifying framework for privacy audits based on information-theoretic principles, modeling the audit as a bit transmission problem in a noisy channel. This formulation allows us to derive fundamental limits and develop an audit approach that yields tight privacy lower bounds for various DP protocols. Second, leveraging this framework, we demystify the method of \textit{privacy audit by one run}, identifying the conditions under which single-run audits are feasible or infeasible. Our analysis provides general guidelines for conducting privacy audits and offers deeper insights into the privacy audit.

Finally, through experiments, we demonstrate that our approach produces tighter privacy lower bounds on common differentially private mechanisms while requiring significantly fewer observations. We also provide a case study illustrating that our method successfully detects privacy violations in flawed implementations of private algorithms.

\end{abstract}

% \Keywords{Node-level Privacy; Graph Neural Networks; Differential Privacy}

% \thispagestyle{plain}
% \maketitle
% \thispagestyle{plain}
% \vspace{-0.4cm}

\section{Introduction}

Safeguarding data privacy has become increasingly important in machine learning tasks, particularly more so when large language models (LLMs) are demanding more data than the whole Internet \cite{not_enough_data}. Among all privacy-enhancing technologies, differential privacy (DP) \cite{dwork2014algorithmic,dwork2009complexity,xiang2024preserving,abadi2016deep,balle2018privacy,hu2023privacy,zhou2024ppml,xiang2023practical} has emerged as a leading paradigm to address these concerns with rigorous mathematical guarantees.
DP ensures that the inclusion or exclusion of a single data point has a minimal impact on the outcome produced by the private algorithm.

Despite its theoretical rigor, implementing differential privacy in practice remains a significant challenge; unintended error often creeps into the realizations. For instance, the sparse vector technique (SVT), a famous differential privacy protocol, has seen erroneous (not private as claimed) applications in some work \cite{zhang2016privtree, chen2015privacy} even though the mathematical proof is given. For another example,  a refinement \cite{stevens2022backpropagation} of the DP-SGD protocol~\cite{bassily2014private, song2013stochastic, abadi2016deep,hu2024differentially,xiang2024preserving,dingrevisiting} that claims to achieve surprisingly strong performance has been proven to suffer from incorrect analysis. Errors are also seen in the implementation phase, where a random seed problem \cite{PRNG_key_reuse}, or a floating-point vulnerability \cite{mironov2012significance} could undermine the integrity of the privacy protocol.

For any real-world application of DP, a straightforward countermeasure is to check the proposed private protocol's analysis or to go through the implemented code line by line. However, this is often cumbersome and also susceptible to errors. These considerations motivate \textit{privacy audit} \cite{nasr2021adversary, kairouz2015composition, jagielski2020auditing}, an empirical approach to measure the privacy provided by differentially private algorithms. It does not check/verify a targeted private algorithm in its detailed implementation; instead, it often involves simulating a \textit{distinguishing game} where an adversary attempts to identify which of two adjacent databases was used as input to run the private algorithm. Intuitively, the targeted algorithm is suggested to be not as private as claimed once accurate identification is achieved.

One standing disadvantage is that such a distinguishing game is usually required to be repeated thousands of times, incurring thousands of times of running the targeted private algorithm itself, because the final probabilistic claim for privacy requires a substantial number of observations to reach non-trivial confidence \cite{nasr2023tight,steinke2022composition,bichsel2021dp,bichsel2018dp}.
This makes it infeasible when running the private algorithm is expensive. Recently, Steinke et al. \cite{steinke2023privacy} propose an audit technique that requires the algorithm (DP-SGD) to run only once while also giving meaningful claims about the empirical privacy level of the targeted algorithm. The critical operation is to perform membership inference \cite{shokri2017membership} on multiple data examples simultaneously based on the result of one run \cite{steinke2023privacy} of the targeted private algorithm. In terms of efficiency, such a \textit{privacy-audit-by-one-run} technique is a substantial
improvement to previous \textit{privacy-audit-by-multiple-run} approaches.

\vspace{0.1cm}
\notbf{Motivation.} The audit story does not end here. We notice several problems worthy of deeper investigation:
\begin{enumerate}
    \item The final empirical privacy claim (known as \textit{privacy lower bound}) by \cite{steinke2023privacy} is not tight in general, e.g., when auditing probably the most well-known Gaussian mechanism, \cite{steinke2023privacy} does not give tight results, even after parameter setups are carefully tweaked and extensively tried; it is unclear how such phenomenon happens. Is such limitation inevitable? Operating under our auditing framework, we can overcome such difficulties, i.e., we can achieve tight results even more efficiently.
    \item To the appealing goal of privacy audit by {one run}, it is also unclear when it is possible/impossible to transfer such a method to other differentially private protocols; are there any helpful universal guidelines for implementing privacy audit by only one run of the targeted private algorithm?  By our unifying language, we provide a bias-variance argument, highlighting when audit-by-one-run is possible or impossible, providing guidance on how to better leverage such a technique.
\end{enumerate}
% 1)The final empirical privacy claim (known as privacy lower bound) is not tight in general, e.g., when auditing probably the most well-known Gaussian mechanism, \cite{steinke2023privacy} does not give tight result, even parameter setups are carefully tweaked; it is unclear how such phenomenon happens. Is such limitation inevitable?  2) to the goal of privacy-audit-by-one-run,
% it is also unclear when it is possible/impossible to transfer such a method to other differentially private protocols; are there any useful universal guidelines for implementing privacy audit by only one run of the targeted private algorithm?
\vspace{-0.1cm}
\notbf{Contribution.} Answering such questions requires a deeper understanding of the privacy audit itself, which is the overall goal we aim to achieve in this paper. Our contribution can be summarized in the following two main parts.

\textbf{1) A unifying language for privacy audit and improved audit method.} We model the privacy audit problem as bits transmission under an information-theoretic context. Behind such treatment is the observation that if some algorithm $\mathcal{M}$ is DP, it ensures indistinguishability between output due to some adjacent input dataset $X,X'$; determining whether it is $X$ or $X'$ (0 or 1) is no difference from recovering \textbf{one bit} of information. Roughly, in the distinguishing game, one chooses $X$ or $X'$ as input to $\mathcal{M}$, and the adversary guesses which one was chosen based on the output of $\mathcal{M}$. In analogy, it coincides with the scenario where a sender aims to communicate one bit of information to a receiver through a noisy channel, corresponding to the execution of $\mathcal{M}$.

Based on such treatment, we study the behavior of such modeling, deriving fundamental limits for bits transmission using the language of information theory. We then leverage those results to design our audit principle. At a high level, if the bits transmission can achieve very low bit error, we can claim (with confidence specification) that $\mathcal{M}$ is not private as promised. 

Except for some carefully designed regulations, our modeling abstracts from the details of how auditing $\mathcal{M}$ is carried out, meaning that our framework can handle both cases of privacy-audit-by-multiple-run and privacy-audit-by-one-run. Practitioners can freely arrange their audit tasks according to regulations, and a privacy lower bound can be derived effortlessly based on our framework. 

\textbf{2) On the tightness of privacy audit and (im)possibilities of privacy audit by one run.} Relying on our ``privacy audit as bits transmission'' modeling, we can answer the two previously raised questions. We show that by modeling $\mathcal{M}$ via $f$-DP \cite{dong2019gaussian} (a DP formulation based on hypothesis testing), we can get tight audit results across various DP protocols; on the other hand, interpretations for why previous work achieves loose audit results are also provided.

Then, based on our framework and theoretical analysis, we 1) demystify privacy-audit-by-one-run and reveal when only one run is (im)possible and 2) provide guidelines on avoiding sub-optimal choices or any other pitfalls for all privacy audit tasks.

In the sequel, targeting the DP-SGD protocol and with the goal of auditing by only \textit{one run}, we carry out experiments of privacy audit on real-world tasks. We show our method achieves better lower bounds than the previous approach to the problem of \textit{privacy audit by one run}. We also give a case study illustrating that our method successfully catches the bug in some ill-implemented private algorithms.

% \vspace{\smgbs}
\section{Background}
% \vspace{\pmgttom}
\subsection{Differential Privacy (DP)} 
\vspace{\pmgttom}
\begin{definition}[Differential Privacy \cite{DBLP:conf/tcc/DworkMNS06}]\label{def:dp}
Let $\mathcal{M}: \mathcal{X}^* \to \mathcal{Y}$ be a randomized algorithm, where $\mathcal{X}^* = \bigcup_{n \ge 0} \mathcal{X}^n$.
    We say $\mathcal{M}$ is $(\varepsilon,\delta)$-differentially private ($(\varepsilon,\delta)$-DP) if, for all $X,X' \in \mathcal{X}^*$ differing only by one element, we have $\forall S \subset \mathcal{Y}$ 
    $$\prob(\mathcal{M}(X) \in S) \le e^\varepsilon \cdot \prob(\mathcal{M}(X') \in S) + \delta.$$
\end{definition}
% \vspace{-0.1cm}
There are two versions of adjacency: 1) for addition/removal, $X'$ has exactly one more data sample than that of $X$; 2) for replacement: $X$ and $X'$ contain the same number of data samples but only differ in exactly one. Our framework in this paper can handle both versions. Post-processing on the output of the DP algorithm is still DP, and the execution of multiple DP algorithms sequentially, known as \textit{composition}, also maintains DP.

\vspace{0.1cm}
\notbf{A functional formulation of DP: $f$-DP.} Using the $(\varepsilon,\delta)$-DP to characterize the privacy of some private algorithm $\mathcal{M}$ has been shown to be lossy \cite{dong2019gaussian}. This is because such a single pair of parameters cannot express the rich nature of the privacy promised by $\mathcal{M}$. In contrast, $f$-DP, based on hypothesis testing formulation, reflects the nature of private mechanisms by a \textit{function} $f$ \cite{zhu2022optimal,dong2019gaussian} rather than a single pair of parameter $(\varepsilon,\delta)$. 

The hypothesis testing setups for $f$-DP is as follows.  Let $\mathcal{Y}$ be the output space of $\mathcal{M}$ taking input one dataset from adjacent datasets $X,X'$, we form the \textit{null}
and \textit{alternative} hypotheses:
% \equienvs
\begin{equation}\label{equ:basic_hypo}
% \equsize
\begin{aligned}
\mathbf{H_0}:\text{$X$ was the input},\text{\quad}\mathbf{H_1}:\text{$X'$ was the input}. 
\end{aligned}
\end{equation}
% \equienve
For a decision rule $\mathcal{R}:\mathcal{Y}\rightarrow \{0,1\}$ for such hypothesis testing setups, two types of errors stand out: 
\begin{itemize}
    \item Type I error or false positive rate $\alpha=\pr{}{\mathcal{R}(y)=1|\mathbf{H_0}}$, i.e., the probability of rejecting $\mathbf{H_0}$ while $\mathbf{H_0}$ is true; 
    \item Type II error or false negative rate $\beta=\pr{}{\mathcal{R}(y)=0|\mathbf{H_1}}$, i.e., the probability of rejecting $\mathbf{H_1}$ while $\mathbf{H_1}$ is true.
\end{itemize}
It is inevitable to make trade-offs between $\alpha$ and $\beta$; what is interesting is the best $\beta$ one can achieve for fixed $\alpha$. This is related to the following definition.

\begin{definition}[Trade-off function \cite{dong2019gaussian}]
    For a hypothesis testing problem over two distributions $P,P'$, define the trade-off function as:
    % \equienvs
    \begin{equation}\nonumber
        T_{P,P'}(\alpha)=\inf_{\mathcal{R}}\{\beta_{\mathcal{R}}:\alpha_\mathcal{R}\leq \alpha\}
    \end{equation}
    % \equienve
    where decision rule $\mathcal{R}$ takes input a sample from $P$ or $P'$ and decides which distribution produced that sample. The infimum is taken over all decision rule $\mathcal{R}$.
\end{definition}

The trade-off function quantifies the best one can do in a hypothesis-testing problem. 
The optimal $\beta$ is achieved via the likelihood ratio test, which is also known as the fundamental \textit{Neyman–Pearson lemma} \cite{neyman1933ix} (please refer to Appendix \ref{app:np_lemma}). For function $f$ and $g$, we denote 
\begin{equation}\nonumber
    \text{$g\geq f$ if $g(x)\geq f(x), \forall x\in[0,1]$}.
\end{equation}
\begin{definition}[$f$-DP \cite{dong2019gaussian}]
    Let $f:[0,1]\rightarrow [0,1]$ be a trade-off function. A mechanism $\mathcal{M}$ is f-DP if 
    % \equienvs
    \begin{equation}\nonumber
        T_{\mathcal{M}(X),\mathcal{M}(X')}\geq f
    \end{equation}
    % \equienve
    holds for all adjacent dataset $X, X'$
\end{definition}
$f$-DP formulation quantifies the indistinguishability between the output of $\mathcal{M}$ due to $X$ or $X'$ by a function, much more expressive than what a single pair of $(\varepsilon,\delta)$ tells. In fact, $f$-DP is a generalization of $(\varepsilon,\delta)$-DP \cite{dong2019gaussian,wasserman2010statistical}: $\mathcal{M}$ is $(\varepsilon,\delta)$-DP equals to  $\mathcal{M}$ is $f_{\varepsilon,\delta}$-DP where the trade-off function $f_{\varepsilon,\delta}$ is 
$$ f_{\varepsilon,\delta}(x)=\max{(0,1-\delta-\mathrm{e}^\varepsilon x,\mathrm{e}^{-\varepsilon}(1-\delta-x))}$$

We also have a useful family of trade-off functions parameterized by $\mu$ as follows.
\begin{definition}[$\mu$-Gaussian DP ($\mu$-GDP) \cite{dong2019gaussian}]\label{def:gdp}
    The trade-off function of distinguishing $\mathcal{N}(0,1)$ from $\mathcal{N}(\mu,1)$ is
    {$$ G_\mu(x)=T_{\mathcal{N}(0,1),\mathcal{N}(\mu,1)}(x)=\Phi(\Phi^{-1}(1-x)-\mu),$$}where $\Phi$ be the c.d.f. of standard normal distribution. A private mechanism $\mathcal{M}$ satisfies $\mu$-GDP if it is $G_\mu$-DP

\end{definition}

\vspace{\pmgttom}
\subsection{Privacy Audit}\label{sec:audit_intro}

An equivalent object to the trade-off function is the ``testing region'' for some algorithm satisfying $(\varepsilon,\delta)$-DP by the following theorem.

\begin{theorem}[$(\varepsilon,\delta)$-DP's testing region \cite{kairouz2015composition}]\label{thm:hp_dp} For any $\varepsilon>0$ and $\delta\in[0,1]$, a mechanism $\mathcal{M}$ is $(\varepsilon,\delta)$-DP if and only if
% \equienvs
\begin{equation}\label{equ:privacy_region}
\begin{aligned}
    & \alpha+\mathrm{e}^\varepsilon \beta\geq 1-\delta, \text{\quad} \beta+\mathrm{e}^\varepsilon \alpha\geq 1-\delta
\end{aligned}
\end{equation}
% \equienve
hold for any adjacent dataset $X,X'$ and any decision rule $\mathcal{R}$ in a hypothesis testing problem defined in Equation \eqref{equ:basic_hypo}.
\end{theorem}

Theorem \ref{thm:hp_dp} bounds the testing region for any decision rule $\mathcal{R}$ if $\mathcal{M}$ is indeed differentially private. The basic principle of privacy audit is to output a privacy lower bound by contraposition of Theorem \ref{thm:hp_dp}, i.e., if some achievable $\alpha,\beta$ falls out of the region defined by $(\varepsilon,\delta)$, it suggests that $\mathcal{M}$ is not $(\varepsilon,\delta)$-DP. Roughly, the general procedure to produce a privacy lower bound is as follows.

After simulating the distinguishing game $n$ times, for each simulation, the adversary makes a binary guess about which of two adjacent datasets was used. This gives the result of many pairs 
\begin{equation}\label{equ:guess_pairs}
    \{(b_g,b_t)_i:i\in[n]\}
\end{equation}
where $b_g$ is the guessed result and $b_t$ is the true secret for one simulation.

Under the empirical approach, we can derive confidence intervals for false positive rate $\alpha\in(\alpha_l,\alpha_r)$ and false negative rate $\beta\in(\beta_l,\beta_r)$ at some confidence level $\gamma$, based on $\{(b_g,b_t)_i:i\in[n]\}$ obtained. This is usually done by the Clopper-Pearson method \cite{clopper1934use} such that $\alpha$ and $\beta$ are modeled as unknown success probabilities
of two binomial distributions. And this leads to the privacy lower bound for the true privacy parameter $\varepsilon_T$ at fixed $\delta$ according to Equation \eqref{equ:privacy_region}.
% \equienvs
\begin{equation}\label{equ:lower_bound}
% \equsize
\begin{aligned}
    \varepsilon_T\geq\varepsilon_L=\max
    \{\log \frac{1-\delta-\alpha_r}{\beta_r}, \log \frac{1-\delta-\beta_r}{\alpha_r}, 0\}
\end{aligned}
\end{equation}
% \equienve
Note that the privacy claim of the private algorithm reports an upper bound $\varepsilon_U\geq\varepsilon_T$ for some fixed $\delta$.

% \vspace{\pmgttom}
\subsection{Related Work}
\notbf{The whole picture of detecting privacy violations.} On detecting privacy violations in the implementation of some differentially private algorithms, there are roughly two mainstream of work that rely on different techniques: 1) using formal verification methods to prove or disprove programs of DP algorithm \cite{wang2020checkdp, zhang2017lightdp,barthe2020deciding,farina2020coupled}; 2) generate refutation for targeted DP algorithms \cite{bichsel2021dp,ding2018detecting, nasr2021adversary,jagielski2020auditing, steinke2023privacy} based on statistical estimation. 

The former often suffers from issues including being not applicable \cite{wang2020checkdp} to $(\varepsilon,\delta)$-DP, Renyi-DP \cite{mironov2017renyi} or another advanced DP application called private selection \cite{liu2019private}; some work also requires necessary manual design assistance \cite{zhang2017lightdp}. Another limitation is that those works cannot handle complex cases where the DP algorithm is part of some larger program \cite{wang2020checkdp}.

In contrast, the latter technique, based on statistical estimation, is free from such issues. Therefore, it is more generalizable, although it may incur the heavy computational overhead of running the targeted DP algorithm thousands of times. Our work also falls under the latter category.

\notbf{The statistic-estimation category.} Privacy audit in this line of work can be framed as aiming to produce a privacy lower bound for the privacy parameter of the targeted algorithm. Certain earlier studies \cite{ding2018detecting,bichsel2021dp, bichsel2018dp,lokna2023group} focus on generating privacy lower bound for some light-weight private protocols, including the Laplace mechanism or sparse vector techniques where running the targeted private algorithm is not a significant computational issue.

Privacy audit in machine learning tasks mainly focuses on investigating the theoretical versus practical privacy guarantees of the DP-SGD protocol \cite{nasr2021adversary, kairouz2015composition, jagielski2020auditing}; one notable work is that Nasr et al. \cite{nasr2021adversary} show that the theoretical privacy analysis for DP-SGD is indeed tight. 

Privacy audit in machine learning benefits from the following lines of related work:
\begin{itemize}
    \item \textbf{Better membership inference}. Under the context of privacy audit, some form of membership inference \cite{shokri2017membership,carlini2022membership} needs to be instantiated in the distinguishing game. For example, Jagielski et al. \cite{jagielski2020auditing} design worst-case data examples, a.k.a. ``\textit{canaries}'' in literature, to form better membership inference, which leads to better privacy lower bound. This line of work aims to produce more informative $\{(b_g,b_t)_i:i\in[n]\}$ (Equation \eqref{equ:guess_pairs}) results of the distinguishing game.

    \item \textbf{Better estimation.} Work in this line aims to perform better statistical analysis over derived $\{(b_g,b_t)_i:i\in[n]\}$ (Equation \eqref{equ:guess_pairs}) results of the distinguishing game, and hence is independent of work on membership inference. For example, Log-Katz confidence interval \cite{lu2022general} and advanced techniques based on Bayesian estimation \cite{zanella2023bayesian} have been proposed to improve the final derived lower bound.
\end{itemize}

\notbf{More efficient privacy audit.} To address the possible computational issue of running the targeted private algorithm many times, improvements have been made on the ``meta-level'': arranging the membership inference and estimation to achieve auditing by fewer runs. 

For instance, Nasr et al. \cite{nasr2023tight} leverage the iterative structure of DP-SGD to perform the overall empirical privacy of DP-SGD; Andrew et al. \cite{andrew2023one} insert random canaries into the input to Gaussian mechanism and measure the privacy based on the result of recovering those random canaries \textit{simultaneously}. Such heuristic of making multiple membership inferences per run of the targeted algorithm has also been used in previous works \cite{zanella2023bayesian,malek2021antipodes} to improve the efficiency of privacy audit; however, such practices are without theoretical rigor; the final estimated privacy lower bound is also considered not faithful \cite{zanella2023bayesian} because membership inferences are performed on data examples not belonging to independent runs, which invalidates current false positive/false negative rate estimation techniques.

Such a problem is further studied by a recent work by Steinke et al. \cite{steinke2023privacy} with theoretical justification. Steinke et al. \cite{steinke2023privacy} also propose an audit method that can derive the final privacy lower bound while requiring the targeted DP-SGD protocol to run only once.

\vspace{\pmgttom}
\subsection{Problem Statement and Motivation}
To briefly describe the approach by Steinke et al. \cite{steinke2023privacy}, 1) first, $n$ contrived data examples (canaries) are decided to be included or not included in the training based on $n$ independent coin flips; 2) second, perform membership inference on those $n$ data examples based on the output of only one run of the targeted algorithm (DP-SGD); 3) finally, privacy lower bound is formed based on the accuracy of those membership inferences.

\notbf{Problems and challenges.} The central contribution made by Steinke et al. \cite{steinke2023privacy} is to validate the operation: performs membership inferences on multiple data examples not belonging to independent runs. However, significant problems with the privacy audit remain unanswered. 

First, the audit method in \cite{steinke2023privacy} is not tight in general, e.g., it does not give tight privacy lower bound for the Gaussian mechanism, even when all parameters are carefully tweaked. Why does this happen, and can it be improved? Second, it is unclear how to transfer such a method to audit problems on DP algorithms other than the DP-SGD protocol. It is beneficial to have some principles to follow.

We also identify another question critical to the audit problem: since the targeted private algorithm is run only once and inserting more canaries becomes a cheap operation, if we ignore other considerations but only focus on deriving better privacy lower bounds, can we arbitrarily increase $n$ to get arbitrarily high confidence for the lower bound estimation?

\notbf{Remark.} To our knowledge, Steinke et al. \cite{steinke2023privacy} are the first to provide privacy lower bound based on only one run of the targeted algorithm with theoretical rigor instead of heuristics. At the time of submission, their method is state-of-the-art. We also note another work \cite{mahloujifar2024auditing} that improves over \cite{steinke2023privacy} based on \cite{steinke2023privacy}'s analysis, reaching slightly better audit results. However, \cite{mahloujifar2024auditing}'s results are still not tight (in contrast, we reach tight results); more importantly, the above raised two problems still remain unanswered. In our experiment, we only compare with \cite{steinke2023privacy} as it suffices to show our contribution related to the above two problems.

\begin{figure*}[!t] 
    \centering
    % \subfloat[]
    {
    \includegraphics[width=.6\linewidth]{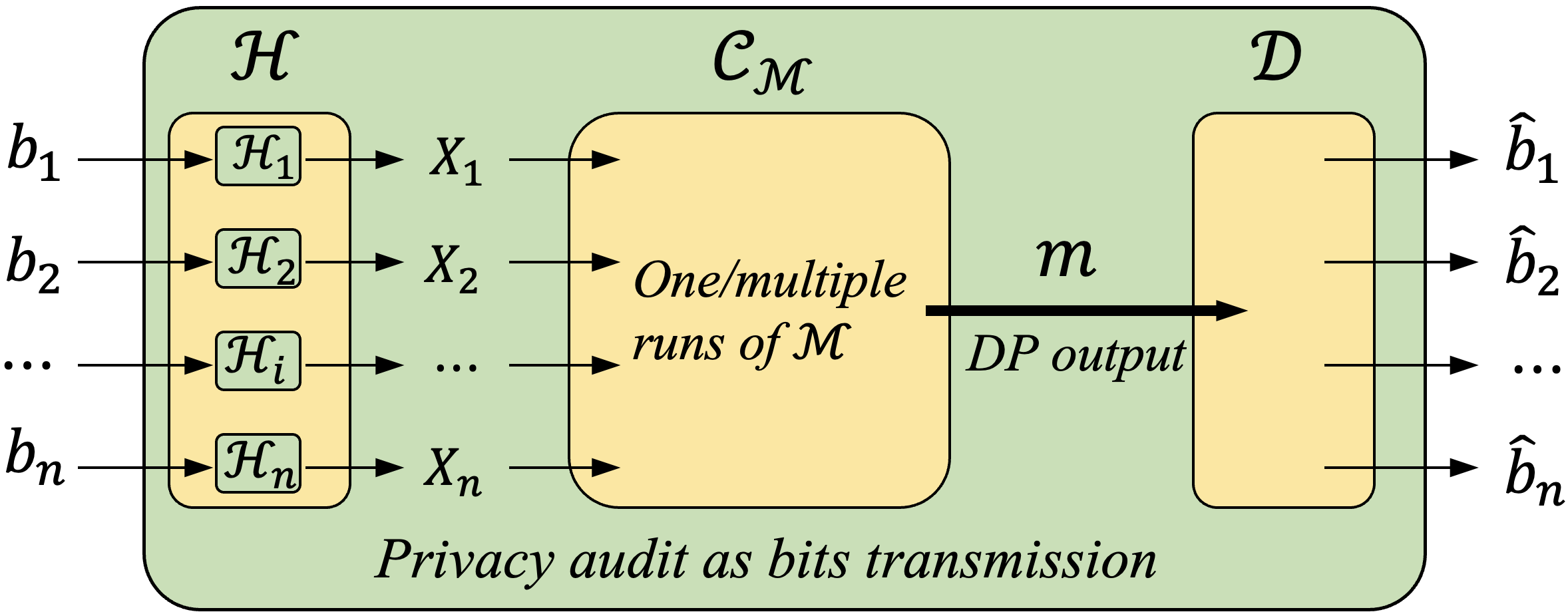}
    }
    \vspace{\fmgftoc}
    \caption{
    The universal framework for privacy audit. Each membership inference corresponds to recovering a bit. Execution of the targeted private algorithm $\mathcal{M}$ corresponds to the usage of a noisy channel for bits transmission. $\mathcal{C}_\mathcal{M}$ is the noisy channel where execution of $\mathcal{M}$ happens, and $\mathcal{D}$ is where the membership inference is launched.  $\mathcal{H}$ is the dataset generator and $m$ is what can be observed by the adversary. 
    }
    \label{fig:channel_modeling} 
    \vspace{\fmgctom}
\end{figure*}

\vspace{\smgbs}
\section{Method}
\vspace{\pmgttom}

\notbf{Method intuition.} Addressing the above questions requires understanding the principle of privacy audit. In this section, we formulate the privacy audit problem as a \textit{bits transmission} framework to serve such a goal. This builds on the observations that, in the membership inference, one of two adjacent datasets needs to be decided as the original input, and this is just equivalent to recovering one bit of information.

Based on our design framework, executing the targeted private algorithm $\mathcal{M}$ is modeled as a use of a noisy channel. If $\mathcal{M}$ is indeed DP, the channel will be noisy enough to prevent reliable bits transmission; therefore, if we can recover bits with low error, we can derive a privacy lower bound for $\mathcal{M}$.

\notbf{Overview.} We use the language of information theory to make such intuition precise. With such an analytical tool, 1) we derive an improved (tight) privacy audit method, which works for both cases of privacy-audit-by-one-run and privacy-audit-by-multiple-run; 2) we also analyze the fundamental limits of privacy audit, which tells us when the appealing audit-by-one goal is possible or impossible.

\vspace{\pmgttom}
\subsection{Universal Framework for Privacy Audit}\label{sec:framework}

\begin{definition}[Privacy audit as bits transmission]\label{def:channel_coding}
    The universal bit transmission framework $(n,p,\mathcal{H},\mathcal{C}_{\mathcal{M}},\mathcal{D})$ for privacy audit models a problem of $n$ bits information transmitting through a noisy channel where
    \begin{enumerate}
        \item $b\in \{0,1\}^n$ is $n$-dimension binary vector where the $i$-th coordinate $b_i$ of $b$ is \underline{independently} sampled from a Bernoulli distribution $\mathbf{Bernoulli}(p)$ $\forall i\in\{1,2,\cdots,n\}$;
        
        \item Dataset generator $\mathcal{H}:\{0,1\}^n\rightarrow \mathcal{X}^n$ outputs an audit dataset $X_A=\{X_i=\mathcal{H}_i(b_i):\forall i \in [n]\}\in\mathcal{X}^n$ where each data sample inside $X_A$ only depends on the corresponding bit of $\mathcal{H}$'s input.
        
        \item The noisy channel $\mathcal{C}_{\mathcal{M}}:\mathcal{X}^n\rightarrow \mathcal{I}$ outputs message $m=\mathcal{C}_{\mathcal{M}}(X_A)\in \mathcal{I}$. $\mathcal{C}_{\mathcal{M}}$ contains built-in information $\mathcal{M}:\mathcal{X}^*\rightarrow \mathcal{Y}$, the DP algorithm we want to audit. $\mathcal{C}_{\mathcal{M}}$ may also contain information of $\mathcal{M}$'s original (training) dataset that is independent of input bits $b$.
        
        \item Decoder $\mathcal{D}:\mathcal{I}\rightarrow \{0,1\}^n$ tries to recover information bits $b$ but actually output $\hat{b}=\mathcal{D}(m)=\{b_i:\forall i\in[n]\}\in \{0,1\}^n$, which might result in errors.
    \end{enumerate}
    \notbf{Regulation:} to comply with privacy audit, it is by design that 1) each data sample inside $X_A$ must only be associated with exactly \underline{\textbf{one}} run of $\mathcal{M}$; 2) message $m=\mathcal{C}_{\mathcal{M}}(X_A)$ must only be formed based on $\mathcal{M}$'s output, i.e., $m$ is differentially private.
\end{definition}

\begin{algorithm}[!ht]
\caption{
$\mathcal{C}_{\mathcal{M}}$ for privacy audit by multiple runs
}\label{alg:channel_coding_for_multiple_run}

\begin{algorithmic}[1]
% \equsize
\renewcommand{\algorithmicrequire}{\textbf{Input:}}
\renewcommand{\algorithmicensure}{\textbf{Output:}}

\Require {$X_A$}
\State $m\gets [ ]$
\For{$i=1,\cdots,n$} %\textbf{in  parallel}
\State Get original training dataset $X_T$
\State $\triangleright$ $X_T$ can be different at each iteration
\State $y_i\gets \mathcal{M}(\{X_A[i]\}\cup X_T)$ \Comment{One run of $\mathcal{M}$}
\State $m.append(y_i)$
\EndFor
\Ensure $m$

\end{algorithmic}
\end{algorithm}

\notbf{Interpretation.} To connect to previous privacy audit terminologies, in our $(n,p,\mathcal{H}, \mathcal{C}_{\mathcal{M}},\mathcal{D})$ framework, the dataset generator $\mathcal{H}$ models the how the canary data examples are formed. The noisy channel $\mathcal{C}_\mathcal{M}$ models the execution of $\mathcal{M}$; being ``noisy'' corresponds to the fact that $\mathcal{M}$ hides the evidence of $X_i$'s participation if $\mathcal{M}$ is indeed DP \cite{DBLP:conf/sp/TschantzSD20}, which makes it hard to recover $b_i$. The decoder $\mathcal{D}$ is where the membership inference happens, such that a binary decision must be made for each input bit.

\vspace{0.1cm}
\notbf{Handling privacy audit by multiple runs.} For previous privacy audit work \cite{nasr2021adversary,jagielski2020auditing,zanella2023bayesian} falling under the category of \text{privacy audit by multiple runs}, these work can be framed by our $(n,p,\mathcal{H},\mathcal{C}_{\mathcal{M}},\mathcal{D})$ framework as follows.

The critical part is what happens inside the noisy channel $\mathcal{C}_{\mathcal{M}}$, for \text{privacy audit by multiple runs}, we demonstrate $\mathcal{C}_{\mathcal{M}}$ as shown in Algorithm \ref{alg:channel_coding_for_multiple_run}. In this line of work, each membership inference is associated with one run of $\mathcal{M}$; here the transmission 
$$b_i\rightarrow X_i \rightarrow y_i=m[i] \rightarrow \hat{b}_i$$
is equivalent to derive a $(b_i,\hat{b}_i)=(b_t,b_g)$ pair as defined in Equation \eqref{equ:guess_pairs}. Note that all $n$ runs of $\mathcal{M}$ are mutually independent.

\begin{algorithm}[!ht]
\caption{
$\mathcal{C}_{\mathcal{M}}$ for privacy audit by one run
}\label{alg:channel_coding_for_one_run}

\begin{algorithmic}[1]
% \equsize
\renewcommand{\algorithmicrequire}{\textbf{Input:}}
\renewcommand{\algorithmicensure}{\textbf{Output:}}
\Require {$X_A$}

\State Get original training dataset $X_T$
\State $y\gets \mathcal{M}(X_A\cup X_T)$ \Comment{One run of $\mathcal{M}$}
\State $m\gets y$

\Ensure $m$
\end{algorithmic}
\end{algorithm}

\vspace{0.1cm}
\notbf{Handling privacy audit by one run.} For previous audit work \cite{steinke2023privacy} falling under privacy audit by one run, such work can be framed by our $(n,p,\mathcal{H},\mathcal{C}_{\mathcal{M}},\mathcal{D})$ framework, shown in Algorithm \ref{alg:channel_coding_for_one_run}. The critical part is that inside $\mathcal{C}_\mathcal{M}$, all generated data examples (corresponding to canaries) are all fed into the input dataset of $\mathcal{M}$ and $\mathcal{M}$ runs on them together only once. In this case, bit transmission may not necessarily be independent, i.e., there might be \textit{interference} between them.

\vspace{0.1cm}
\notbf{Handling two versions of adjacency.} Our framework applies to audit tasks for both versions of the adjacency definition of DP. If the $\mathcal{M}$ is DP with respect to addition/removal, we require $X_i$ to be either a real canary data example or a \textit{null} object that contributes nothing to the execution of $\mathcal{M}$, based on $b_i$. This null setup models the case where the canary is not included in $\mathcal{M}$'s. If the $\mathcal{M}$ is DP with respect to replacement, $X_i$ can be any different data examples depending on $b_i$.

\vspace{0.1cm}
\notbf{Basic considerations.} The final goal is to estimate the privacy lower bound of $\mathcal{M}$ based on $b=\{b_i:\forall i\in[n]\}$ and $\hat{b}=\{\hat{b}_i:\forall i\in[n]\}$. By basic design principles, as we can always let $\hat{b}$ be bad guesses (e.g., just make $\hat{b}$ random), it is pivotal for $\hat{b}$ to recover $b$ as accurate as possible, which allows to conclude more informative assertion about $\mathcal{M}$'s privacy (deriving stronger privacy lower bound).

\begin{table}[!ht] 
\equsize
% \vspace{-0.1cm}
\centering
% \begin{subtable}[h]{\textwidth}
% \raggedleft
% \centering
\resizebox{0.99\columnwidth}{!}{
% \left
\begin{tabular}{cl}
% \toprule
\toprule

Notation & Meaning\\
\midrule
$B$ & Random input bits sampled from $\mathbf{Bernoulli}(p)^n$\\
$B_i$ & Marginal distribution of $i$-th coordinate of $B$\\
$\hat{B}$ & Random recovered bits\\
$\hat{B}_i$ & Marginal distribution of $i$-th coordinate of $\hat{B}$\\
\midrule
$M$ & Random variable for $\mathcal{C}_\mathcal{M}$'s output\\
$E_i$ & Bit error random variable defined in Equation \eqref{equ:bit_error_rv}\\
$E$ & Bit error random vector $E=[E_1, \cdots, E_n]$\\
\midrule
\multicolumn{2}{c}{For some random vector $R$}\\
\midrule
$R_{-i}$ & Random vector $R_{-i} = [R_1, \cdots, R_{i-1}, R_{i+1},\cdots R_{n}]$\\
$R_{<i}$ & Random vector $R_{<i} = [R_1, \cdots, R_{i-1}]$\\
\midrule
\multicolumn{2}{c}{\textbf{Lowercase use corresponds to the above's realization}}\\

\bottomrule
% \caption{accuracy}
\end{tabular}

}

\caption{
Notation for random variables used. 
}
\label{tab:notations}
% \vspace{-0.5cm}
\end{table}

\subsection{Information-theoretic Limits}

In this section, we give results demonstrating the fundamental information-theoretic limits under our audit framework $(n,p,\mathcal{H},\mathcal{C}_{\mathcal{M}},\mathcal{D})$. Those limits always hold if complying with our framework, regardless of whether the privacy audit is by one run or multiple run of $\mathcal{M}$.

\notbf{Notation.} We use uppercase letters (e.g., $Z$) to represent a random variable and lowercase letters (e.g., $z$) to denote its realization.
When we need to refer to the distribution of $Z$, we also abuse using the uppercase without ambiguity, e.g.,
\begin{equation}
Z|_{Y=y}    
\end{equation}
denotes the distribution of random variable $Z$ conditioned on random variable $Y=y$. Notations for random variables used are summarised in Table \ref{tab:notations}.

We care about the errors made in bits transmission, which is formally defined in the following.

\begin{definition}\label{def:all_error}
    Define a random variable $E_i$ as follows.
    \begin{equation}\label{equ:bit_error_rv}
        E_i = \left\{\begin{array}{ll}
        1 \textit{,\hspace{.05cm} if \hspace{.1cm}} \hat{B}_i \neq B_i\\
        0 \textit{,\hspace{.05cm} if \hspace{.1cm}} \hat{B}_i = B_i \\
        \end{array}\right. 
    \end{equation}
    I.e., $E_i$ is the random variable indicating whether recovered bit $\hat{B}_i$ is an error. We denote 
    \begin{equation}\label{equ:bit_error}
        p_i^{e}=\pr{}{E_i=1}
    \end{equation} 
    as the \textbf{bit error} for $i$. We also define the \textbf{average} bit error as
    \begin{equation}\label{equ:avg_bit_error}
        p^{e}=\frac{1}{n}\sum_i\pr{}{E_i=1}=\frac{1}{n}\sum_i p_i^e
    \end{equation} 
\end{definition}

We will also formalize the indistinguishability provided by DP algorithm under our framework using $f$-DP. If distribution $P, P'$ possessing some level of indistinguishability and their trade-off function satisfies
\begin{equation}\nonumber
T_{P,P'}\geq f
\end{equation}
we denote this relation between $P, P'$  as 
\begin{equation}\label{equ:f_DP_relation}
P \overset{f-\text{DP}}{\sim} P'.
\end{equation}
It is easy to see that $\mathcal{M}$ is $f$-DP if $\mathcal{M}(X) \overset{f-\text{DP}}{\sim} \mathcal{M}(X')$ for all adjacent $X,X'$.

\vspace{0.1cm}
\notbf{Implication from our framework regulation and differential privacy.} With all the notations set, we are ready to state some facts based on our framework and differential privacy. We have the following property according to the regulation shown in Definition \ref{def:channel_coding}.

\begin{property}[Noisy transmission implied by DP, proof in Appendix \ref{app:proof_bit_fdp_conditioned}]\label{property:bit_fdp_conditioned}
    In our $(n,p,\mathcal{H}, \mathcal{C}_{\mathcal{M}},\mathcal{D})$ framework, if $\mathcal{M}$ is $f$-DP, then $\forall i\in [n], b_{-i}\in\{0,1\}^{n-1}$
    \begin{equation}\label{equ:bit_fdp_conditioned}
    \begin{aligned}
        \hat{B}_i|_{B_i=0,B_{-i}= b_{-i}}\fdp{f} \hat{B}_i|_{B_i=1,B_{-i}= b_{-i}}
    \end{aligned}
    \end{equation}
    Where $[n]=\{1,2,\cdots,n\}$. Equation \eqref{equ:bit_fdp_conditioned}  intuitively says that, conditioned on other input bits being fixed to be $b_{-i}$, even if we flip the $i$-th input bit, its corresponding output bit's distribution will not change much. 
\end{property}

We present a lemma that we later rely on as follows.
\begin{lemma}[Mixture by convex combination only makes it more indistinguishable, proof in Appendix \ref{app:proof_mix_become_harder}]\label{lem:mix_become_harder}
    If $P_i \overset{f-\text{DP}}{\sim} P'_i, \forall i=1,\cdots,n$, then $\forall c_i\in[0,1]$ such that $\sum_{i=1}^n c_i=1$, we have
    \begin{equation}
        \sum_{i=1}^n c_i P_i \overset{f-\text{DP}}{\sim} \sum_{i=1}^n c_i P'_i.
    \end{equation}
    intuitively, the pair of corresponding mixture distributions by convex combination becomes only harder to distinguish than each $P_i, P'_i$ pair.
\end{lemma}

\notbf{Hardness of single bit transmission}. Under our $(n,p,\mathcal{H},\mathcal{C}_{\mathcal{M}},\mathcal{D})$ framework, the marginal distribution for some single output bit also possesses some level of indistinguishability condition on different corresponding input bit.

\begin{figure}[!ht] 
    \centering
    % \subfloat[Histograms]
    {\includegraphics[width=.55\linewidth]{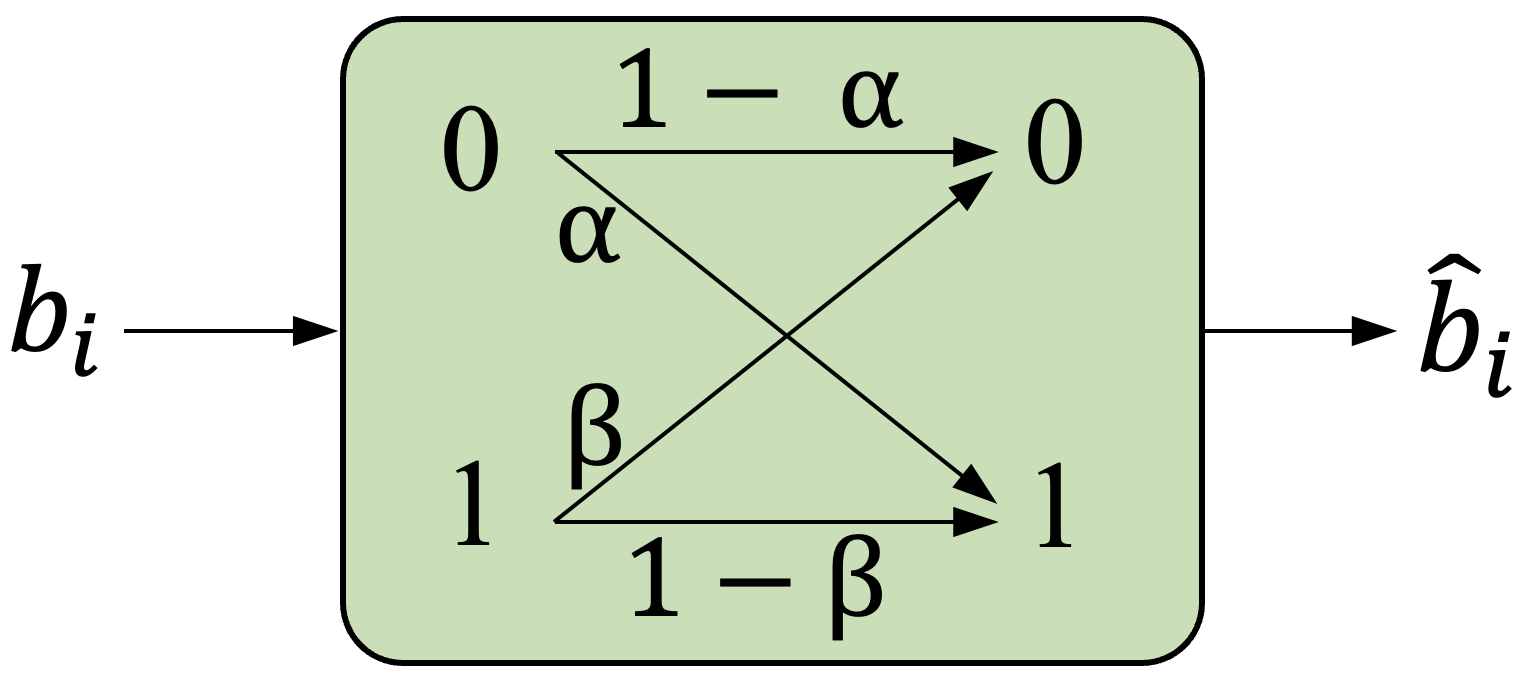}}
    \vspace{\fmgftoc}
    \caption{Single-bit transmission, modeled as a binary channel. If input bit $b_i=0$, the channel flips the bit with probability $\alpha$, corresponding to a false positive rate; if $b_i=1$, the bit is flipped with probability $\beta$, which is the false negative rate. As governed by the trade-off function, $\beta\geq f(\alpha)$ must hold. if $\alpha=\beta$, the above channel is the well-known binary symmetric channel (BSC).}
    \label{fig:bit_binary_channel} 
    % \vspace{-0.3cm}
\end{figure}

\begin{corollary}[Recovering bits is hard, proof in Appendix \ref{app:proof_marginal_bit_recover_hard}]\label{cor:marginal_bit_recover_hard}
    In our $(n,p,\mathcal{H},\mathcal{C}_{\mathcal{M}},\mathcal{D})$ framework, if $\mathcal{M}$ is $f$-DP, then $\forall i\in[n]$, 
    \begin{equation}\label{equ:marginal_bit_recover_hard}
        \hat{B}_i|_{B_i=0}\fdp{f} \hat{B}_i|_{B_i=1}
    \end{equation}
\end{corollary}
We need the help of Lemma \ref{lem:mix_become_harder} to complete the proof provided in Appendix \ref{app:proof_marginal_bit_recover_hard}. Based on this result, we abstract the channel for the transmission of the $i$-th bit in Figure \ref{fig:bit_binary_channel}. We define the null hypothesis as $b_i=0$ and the alternative hypothesis as $b_i=1$. An arbitrary decision rule $\mathcal{R}$ makes a false negative rate $\beta$ as a false positive rate $\alpha$.

Hence, our channel molding tells us recovering the input bits is fundamentally hard due to the noisy channel $\mathcal{C}_\mathcal{M}$. How do we quantify such hardness? We use the information-theoretic quantity: \textit{mutual information}, a central topic in information theory. We use 
\begin{equation}\label{equ:mi}
    \mathrm{MI}(B_i;\hat{B}_i)
\end{equation}
to denote the mutual information between two binary random variables $B_i$ and $\hat{B}_i$. A closely related quantify is the channel capacity at fixed $\alpha,\beta$
\begin{equation}\label{equ:channel_capa}
    \mathbb{C} = \max_{p}\mathrm{MI}(B_i;\hat{B}_i)
\end{equation}
Under the original information-theoretic context, channel capacity is the maximum rate at which we can send information over some noisy channel with a vanishingly low error probability. Under our privacy audit context, it is related to the best bit error that can be achieved, as will be shown later.

Assuming we are using $2$ as the base of logarithm 
for information related quantities throughout this paper, the binary entropy function is 
\begin{equation}\label{equ:binary_entropy}
    \mathrm{h}(x)=-x\cdot \log(x)-(1-x)\cdot\log(1-x)
\end{equation}
where $\mathrm{h}(x):[0,1]\rightarrow [0,1]$ is symmetric around $x=\frac{1}{2}$. Also, define an inverse function 
\begin{equation}\label{equ:inverse_binary_entropy}
\mathrm{h}^{-1}(x):[0,1]\rightarrow[0,\frac{1}{2}]
\end{equation}
corresponding to the inverse of the left half part of $\mathrm{h}(x)$ defined on $x\in [0,\frac{1}{2}]$.
% $\mathrm{h}_l^{-1}$ & inverse function of $\mathrm{h}$ defined over $x\in [0,\frac{1}{2}]$\\

Intuitively, mutual information $\mi(B_i;\hat{B}_i)$ measures the dependence between random bits $B_i$ and $\hat{B}_i$. In our $(n,p,\mathcal{H},\mathcal{C}_{\mathcal{M}},\mathcal{D})$ framework, note that $B_i\sim\mathbf{Bernoulli}(p)$, if $\mi(B_i;\hat{B}_i)$ reaches to its maximum $\mi(B_i;\hat{B}_i)=\mathrm{H}(B_i)=\mathrm{h}(p)$ where $\mathrm{H}(B_i)$ is the entropy of $B_i$, we then have $\hat{B}_i=B_i$, i.e., perfect transmission. 

However, the algorithm $\mathcal{M}$ prevents perfect transmission if it is private. We have the following results in quantifying the fundamental hardness of single-bit transmission.

\begin{figure}[!t] 
    \centering
    % \subfloat[Histograms]
    {\includegraphics[width=.9\linewidth]{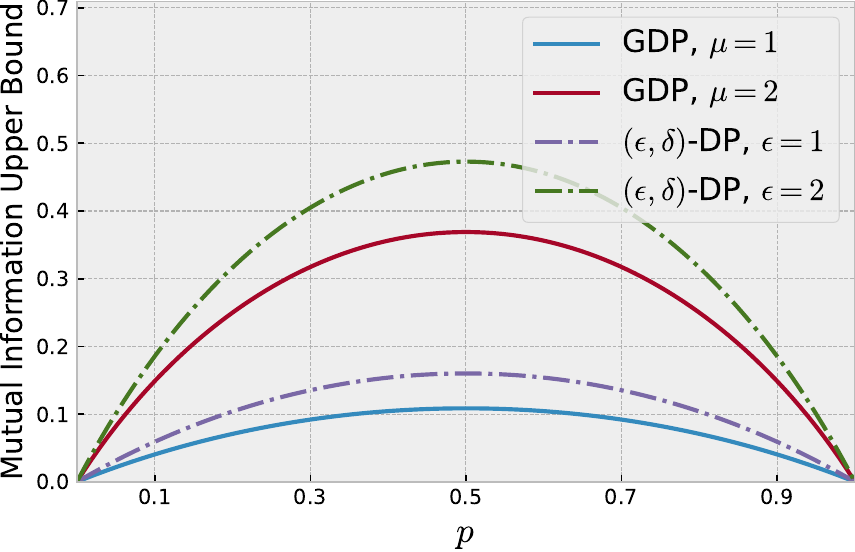}}
    % \vspace{-0.1cm}
    \vspace{-0.3cm}
    \caption{
    The mutual information upper bound $u_f(p)$ for different trade-off functions. $\delta=10^{-5}$ for $(\varepsilon,\delta)$-DP.
    }
    \label{fig:mi_at_diff_p} 
    % \vspace{-0.3cm}
\end{figure}

\begin{theorem}[Mutual information upper bound for bits transmission, proof in Appendix \ref{app:proof_mi_bound_bit_trans}]\label{thm:mi_bound_bit_trans}
    in our $(n,p,\mathcal{H}, \mathcal{C}_{\mathcal{M}},\mathcal{D})$ framework, if $\mathcal{M}$ is $f$-DP, we have $\forall i\in[n]$
    \begin{equation}\label{equ:mi_bound_bit}
        \mi(B_i;\hat{B}_i)\leq \max_{x\in[0,1]} F_f(x,p)\overset{\text{def}}{=}u_f(p)
    \end{equation}
    where 
    \begin{equation}\label{equ:upper_bound_p}
    \begin{aligned}
        F_f(x,p)=&\mathrm{h}(p\cdot f(x)+(1-p)(1-x))-p\cdot\mathrm{h}(f(x))\\
        &-(1-p)\cdot\mathrm{h}(1-x)\\
    \end{aligned}
    \end{equation}
    % where function $u$ takes input a trade-off function $f$ and outputs 
\end{theorem}
Proof of this theorem relies on our channel modeling shown in Figure \ref{fig:bit_binary_channel}. In practice, computing $u_f(p)$ is always numerically stable as all terms are bounded. Figure \ref{fig:mi_at_diff_p} plots the upper bound function $u_f(p)$, and we can see that the more private trade-off function tends to have a smaller upper bound value.

Equation \eqref{equ:upper_bound_p} leads to a somewhat complex form because we allow $p$ to be chosen freely. We can make it concise by setting $p=\frac{1}{2}$, i.e., each bit is independently and uniformly sampled from $\{0,1\}$. This setup corresponds to the balanced prior adopted in previous work \cite{shokri2017membership,nasr2021adversary,salem2023sok}. And we assume $p=\frac{1}{2}$ in the remaining part unless specified otherwise.

Theorem \ref{thm:mi_bound_bit_trans} also tells us that, even if the membership inference does its best, i.e., the false positive rate $\alpha$ and false negative rate $\beta=f(\alpha)$ lands on boundaries defined by the trade-off function, the mutual information still has an upper bound. Consequently, it leads to unavoidable non-trivial error in bits transmission, as stated by the following result.

\begin{theorem}[Bit error lower bound, proof in Appendix \ref{app:proof_bit_error_lower_bound}]\label{thm:bit_error_lower_bound}
    In our $(n,\frac{1}{2},\mathcal{H},\mathcal{C}_{\mathcal{M}},\mathcal{D})$ framework,
    w.o.l.g., assume $p_i^{e}\leq \frac{1}{2}$ (defined in Equation \eqref{equ:bit_error}), because one can always do better than random guessing. If $\mathcal{M}$ is $f$-DP, we have $\forall i\in [n]$
    \begin{equation}\label{equ:bit_error_lower_bound}
        p_i^{e}\geq \mathrm{h}^{-1}(1- u_f(\frac{1}{2}))\overset{\text{def}}{=}p_f^e
    \end{equation}
    Where $\mathrm{h}^{-1}$ is defined in Equation \eqref{equ:inverse_binary_entropy}.
\end{theorem}
Theorem \ref{thm:bit_error_lower_bound} is the basis of our audit method shown in the following section.

\vspace{\pmgttom}
\subsection{Audit Method}

\notbf{Method overview.} Our idea for privacy audit is fairly simple: if $\mathcal{M}$ is some $f$-DP, 1) for each bit transmission, the bit error will not be too small; 2) therefore, if we observe some significantly low bit error, we can conclude (with confidence specification) that $\mathcal{M}$ is not $f$-DP as claim by contraposition. This gives a privacy lower bound.

However, deriving a lower bound requires non-trivial manipulation. Our method in this section quantifies the relationship between bit error and the lower bound that we can conclude.

\notbf{Privacy audit without interference.} We first assume that each bit transmission $b_i\rightarrow \hat{b}_i$ is mutually independent. Under the information-theoretic context, this equals to the fact that there is \textit{no interference} between bits transmission; another equivalent characterization is that the noisy channel $\mathcal{C}_\mathcal{M}$ is a \textit{memoryless} channel.

For the privacy audit task design, whether interference exists between bits transmission can always be controlled. For example, we can always resort to the privacy-audit-by-multiple-run case where interference does not exist. Actually, for privacy audits, ensuring no interference should be viewed as a principle to follow. We will provide this argument in later sections, and we will also provide an analysis for the case where interference exists.

In the case where interference does not exist, the random variable $E_i$ ((defined in Equation \eqref{equ:bit_error_rv} ) is independent of $E_j$, $\forall i,j\in[n]$ and $ i\neq j$. Note that the error probability $\{p_i^e:\forall i\in[n]\}$ (defined in Equation \eqref{equ:bit_error}) may not necessarily be the same. 

\begin{theorem}[Audit principle, proof in Appendix \ref{app:proof_bit_error_dominance_independent}]\label{thm:bit_error_dominance_independent}
      In our $(n,\frac{1}{2},\mathcal{H},\mathcal{C}_{\mathcal{M}},\mathcal{D})$ framework, if $E_i$ (defined in Equation \eqref{equ:bit_error_rv} ) is independent from $E_j$, $\forall i,j\in[n]$ and $ i\neq j$, let $S$ be a $n$-dimension binary vector where each coordinate of $S$ is independently sampled from $\mathbf{Bernoulli}(p_f^e)$ ($p_f^e$ defined in \eqref{equ:bit_error_lower_bound}). We have $\forall a\in[0,1]$  \begin{equation}\label{equ:bit_error_dominance_independent}
         \pr{E}{\frac{1}{n}\sum_i^n E_i\geq a}\geq  \pr{S}{\frac{1}{n}\sum_i^n S_i\geq a}
     \end{equation}
\end{theorem}
This theorem says that it is more likely to see a greater value for the average of results sampled from $\{E_i:\forall i\in[n]\}$ than that of results sampled from $n$ independent Bernoulli tries with probability $p_f^e$. Intuitively, this is because the probability of making an error ($E_i=1$) for each bit is greater than $p_f^e$, which is the best we can do. It allows us to make the following deductions, which also form the basis of our audit method.

\notbf{Confidence interval (CI) construction.} In our $(n,\frac{1}{2},\mathcal{H},\mathcal{C}_{\mathcal{M}},\mathcal{D})$ framework, if $\mathcal{M}$ is $f$-DP, 
\begin{enumerate}
    \item If we really can achieve the ``best''. The bit error random variable becomes $E_i=S_i\sim \mathbf{Bernoulli}(p_f^e), \forall i\in[n]$.;
    \item For pre-defined confidence level $\gamma$ (95\% and 99\% are typical), we can compute $v$ such that 
    \begin{equation}\label{equ:ber_geq_gamma}
    \pr{S}{\frac{1}{n}\sum_i S_i\geq p_f^e-v}=\gamma
    \end{equation}
    \item For some real random variable $E_i$ that we can actually achieve, Theorem \ref{thm:bit_error_dominance_independent} promises that 
    \begin{equation}\label{equ:geq_gamma}
        \pr{}{\frac{1}{n}\sum_i E_i\geq p_f^e-v}\geq\gamma
    \end{equation}
    \item Hence, once we observe an outcome $\bar{e}$ for the random variable of the sample mean $\frac{1}{n}\sum_i E_i$,  Equation \eqref{equ:geq_gamma} give us the confidence interval
    \begin{equation}\label{equ:ci}
        [0, \bar{e}+v]
    \end{equation}
    for $p_f^e$ (the parameter we aim to estimate) with a coverage level always greater than $\gamma$.
\end{enumerate}
The remaining task is how to compute $v$ and we can let $v=v(n,\gamma)$, a function of $n, \gamma$, which can be tackled by the Hoeffding bound as follows

\begin{equation}\label{equ:hoeffding_bound_interval}
    v = \sqrt{\frac{1}{2n}\log\frac{1}{1-\gamma}}
\end{equation}
The derivation for Equation \eqref{equ:hoeffding_bound_interval} is a standard application of Hoeffding's inequality, and it is presented in Appendix \ref{app:hoeffding_bound}. 
% With this result, 

\begin{algorithm}[!t]
\caption{
Advanced CI \hspace{0,1cm} $\mathcal{ACI}(\bar{e},\gamma,n)$
}\label{alg:advanced_CI}

\begin{algorithmic}[1]
% \equsize
\renewcommand{\algorithmicrequire}{\textbf{Input:}}
\renewcommand{\algorithmicensure}{\textbf{Output:}}
\Require {$\bar{e}$, empirical average bit error; $\gamma$ confidence specification; $n$, total number of bits transmission}

\State $p_l\gets0.001$, $p_r\gets\frac{1}{2}$
\State $p_{min}^e\gets\frac{1}{2}(p_l+p_r)$
\State $\triangleright$ $F^{-1}$ is the inverse c.d.f. of a Binomial distribution
\State $v\gets p_{min}^e-\frac{1}{n}F^{-1}\left(1-\gamma,n,p_{min}^e\right)$\label{alg:advanced_CI_v}
\For{$\|p_{min}^e-(\bar{e}+v)\|>0.0001$}
\If{$p_{min}^e>\bar{e}+v$}
\State $p_r^e\gets p_{min}^e$
\Else{}
\State $p_l^e\gets p_{min}^e$
\EndIf
\State $p_{min}^e\gets\frac{1}{2}(p_l+p_r)$
\State $v\gets p_{min}^e-\frac{1}{n}F^{-1}\left(1-\gamma,n,p_{min}^e\right)$
\EndFor

\Ensure $p_{min}^e$
\end{algorithmic}
\end{algorithm}

\vspace{0.1cm}
\notbf{A more sophisticated CI construction method.} Computing $v$ by Equation \eqref{equ:hoeffding_bound_interval} is simple, however, we can do better. For completeness, we also provide another more sophisticated CI construction method in the following, but paying the price for a slightly higher complexity. 

The general idea is to let $v=v(p_f^e,n,\gamma)$, i.e., it also depends on $p_f^e$, and we can derive better results for $v$ by iteration. The high-level intuition and procedure are as follows. After seeing $\bar{e}$ (which we cannot control in estimation), we want $v$ as small as possible because it will lead to better privacy lower bound. Then, we can compute $v$ by Equation \eqref{equ:ber_geq_gamma} based on Binomial distribution; we repeat such process by setting different hypothetical values for $p_f^e\gets p_{min}^e$ until a certain condition is met. The detailed method is provided in Algorithm \ref{alg:advanced_CI}.

\begin{algorithm}[!ht]
\caption{
$f$-DP to $(\varepsilon,\delta)$-DP \cite{dong2019gaussian} \hspace{0.1cm} $\mathcal{ED}(f, \delta)$
}\label{alg:fdp_to_eps_delta}

\begin{algorithmic}[1]
% \equsize
\renewcommand{\algorithmicrequire}{\textbf{Input:}}
\renewcommand{\algorithmicensure}{\textbf{Output:}}

\Require {$f$, trade-off function; $\delta$, privacy parameter}

\State  $\varepsilon \gets \infty$ if $\delta < 1-f(0)$; Return $\infty$
\State Compute $\varepsilon=\inf\{a:f(x)\geq 1-\delta-\mathrm{e}^a x, \forall x\in[0,1]\}$ via binary search

\Ensure $\max\{0, \varepsilon\}$

\end{algorithmic}
\end{algorithm}

\begin{algorithm}[!ht]
\caption{
Privacy lower bound $\mathcal{LB}(\delta, \bar{e},\gamma,n)$
}\label{alg:lb_by_privacy_audit}

\begin{algorithmic}[1]
% \equsize
\renewcommand{\algorithmicrequire}{\textbf{Input:}}
\renewcommand{\algorithmicensure}{\textbf{Output:}}
\Require {$\delta$, privacy parameter; $\bar{e}$, empirical average bit error; $\gamma$ confidence specification; $n$, total number of bits transmission}

% \State $v \gets \sqrt{\frac{1}{2n}\log\frac{1}{1-\gamma}}$\label{alg:lb_by_privacy_audit_v_compute}
\State $p_f^e \gets \sqrt{\frac{1}{2n}\log\frac{1}{1-\gamma}}$ or $p_f^e \gets \mathcal{ACI}(\bar{e},\gamma,n)$ \Comment{Algorithm \ref{alg:advanced_CI}}
\State Compute $f$ s.t. $\mathrm{h}(p_f^e)=1- u_f(\frac{1}{2})$ \Comment{Equation \eqref{equ:bit_error_lower_bound}}
\State $\varepsilon_L \gets\mathcal{ED}(f, \delta)$ \Comment{Algorithm \ref{alg:fdp_to_eps_delta}}

\Ensure $\varepsilon_L$
\end{algorithmic}
\end{algorithm}

\vspace{0.1cm}
\notbf{Deriving the final privacy lower bound.} we can derive the final privacy lower bound after the confidence interval is constructed. The lower bound is in $(\varepsilon,\delta)$-DP formulation, aligning with almost all previous work. The whole process is presented in Algorithm \ref{alg:lb_by_privacy_audit}. Note that in our method, we essentially estimate an upper bound for averaged bit error, which corresponds to an upper bound of $f$-DP, which can be converted into a final privacy lower bound $\varepsilon_L$. The conversion is presented in Algorithm \ref{alg:fdp_to_eps_delta}.

\begin{figure}[!t] 
    \centering
    \subfloat[$\bar{e}=0.2,n=100, \gamma=0.99$]
    {
    \includegraphics[width=.8\linewidth]{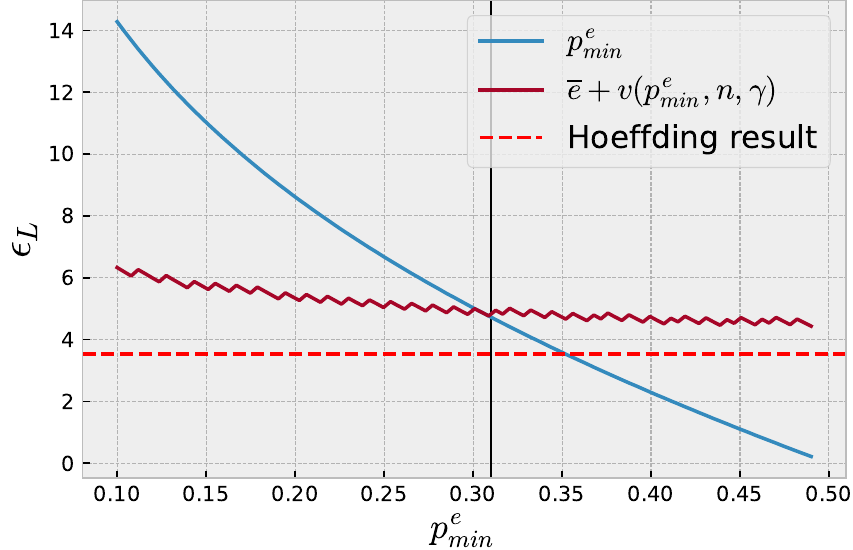}
    }\\
    \subfloat[$\bar{e}=0.2,n=10^{3},\gamma=0.99$]
    {
    \includegraphics[width=.8\linewidth]{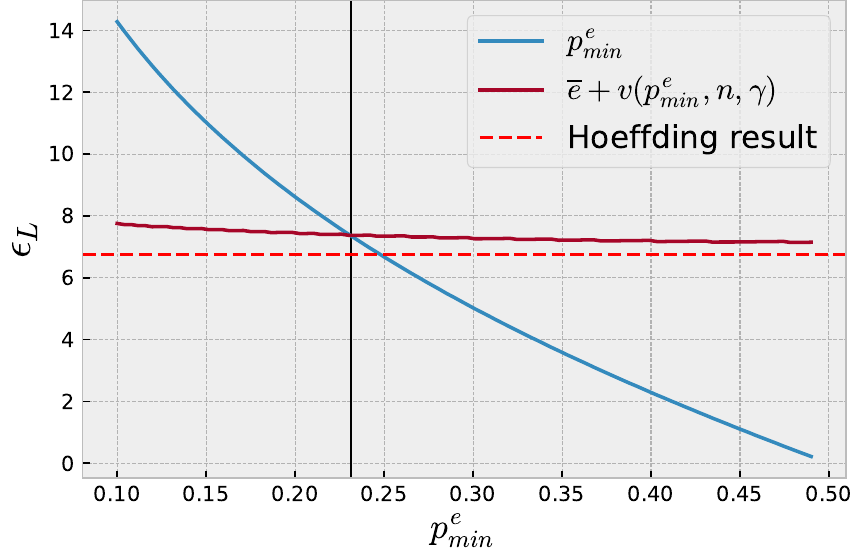}
    }
    \vspace{\fmgftoc}
    \caption{
    Illustration on how our advanced CI method works. The horizontal axis is different $p_{min}^e$ value we assume that we can achieve, and the vertical axis is the corresponding lower bound. Line marked as $p_{min}^e$ corresponds to lower bound derived based on average bit error $p_{min}^e$ can be achieved; the same is to $\bar{e}+v(p_{min}^e,n,\gamma)$. Hoeffding result is the simple CI result shown in Equation \eqref{equ:hoeffding_bound_interval}. Regions on the left of the vertical black line are where we have contradictions. 
    }
    \label{fig:dif_ci_construction} 
    \vspace{\fmgctom}
\end{figure}

\vspace{0.1cm}
\notbf{Understanding the advanced CI method.} After introducing how the final lower bound is derived, we elaborate on how our advanced technique works. Recall that we set different hypothetical values $p_{min}^e$ to $p_f^e$; each time we set a value to $p_f^e$, we are essentially making an assumption on the privacy lower bound. Specifically, if we assume we can achieve average bit error $p_{min}^e$, we are assuming the privacy lower bound derived based on this assumption is at least some value $\varepsilon_L(p_{min}^e)$ where $\varepsilon_L(p_{min}^e)$ is computed by the last two lines in Algorithm \ref{alg:lb_by_privacy_audit} by setting $p_f^e\gets p_{min}^e$. 

Based on assuming that $p_{min}^e$ is the best we can achieve, we can derive the upper bound $\bar{e}+v(p_{min}^e,n,\gamma)$ of the confidence interval. If $\varepsilon_L(\bar{e}+v(p_{min}^e,n,\gamma))<\varepsilon_L(p_{min}^e)$, we have a contradiction, which requires we revise the assumption $p_{min}^e$ until there are no more contradictions. The illustration is provided in Figure \ref{fig:dif_ci_construction}. Our advanced method has a notable advantage over the Hoeffding method when we do not have a large number of observations ($n$ is small).

% \vspace{\pmgttom}
\subsection{Principle for Channel Arrangement}

\begin{figure}[!ht] 
    \centering
    \subfloat[Memoryless channel]
    {
    \includegraphics[width=.47\linewidth]{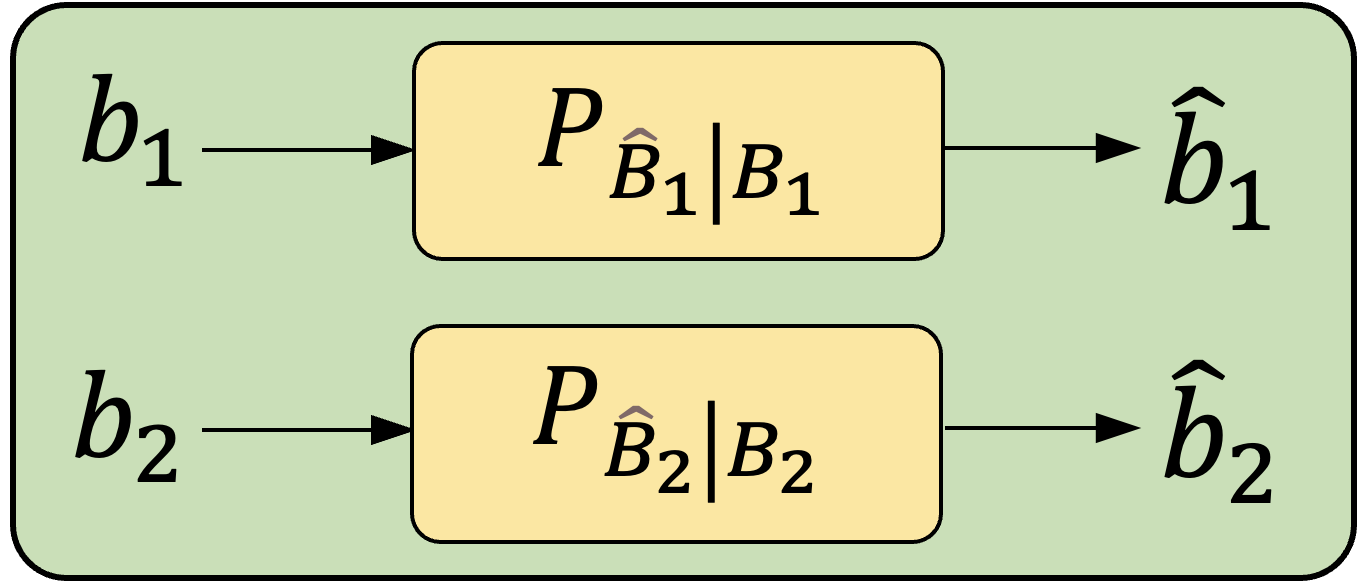}
    }
    \subfloat[Channel with interference]
    {
    \includegraphics[width=.47\linewidth]{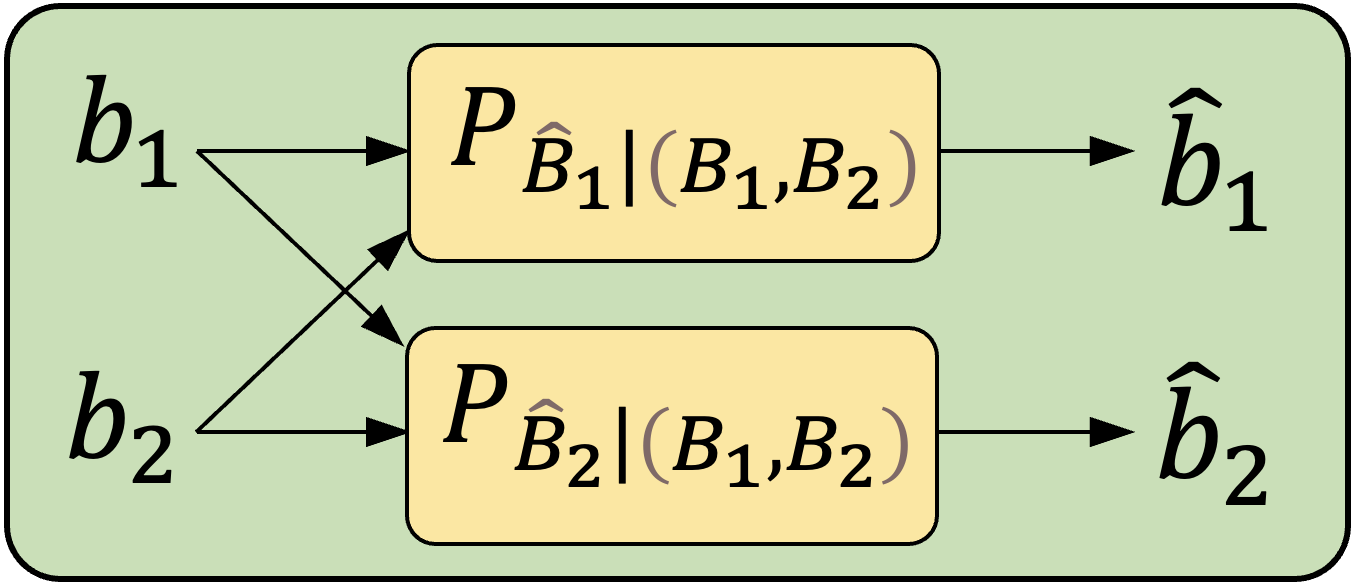}
    }
    \vspace{\fmgftoc}
    \caption{
    Diagram of the case where we have $n=2$ bits of transmission. When there is interference, output bit $\hat{b}_1$ also depends on input $b_2$, but $\hat{b}_1$ is only intended to recover $b_1$.
    }
    \label{fig:two_dif_channel} 
    \vspace{\fmgctom}
\end{figure}

In this section, we will justify why arranging the channel $\mathcal{C}_\mathcal{M}$ to be a memoryless channel for bits transmission is always superior to bits transmission with interference. In our $(n,p,\mathcal{H},\mathcal{C}_{\mathcal{M}},\mathcal{D})$ framework, we will also see the fact that avoiding interference should stand as an unquestioned design principle for all audit tasks.

\notbf{Performance metric and sub-channel arrangement.} To illustrate both cases for memoryless channel arrangement and channel with interference, we show an example in Figure \ref{fig:two_dif_channel} when $n=2$. When there is no interference, we use $\mathbf{P}_{\hat{B}_i|B_i}$ to denote the transition probability matrix or Markov kernel for the channel where $B_i$ and $\hat{B}_i$ are the input and output random variable, respectively. Such Markov kernel models the overall effect due to $\mathcal{H}$, $\mathcal{C}_\mathcal{M}$ and $\mathcal{D}$. In the presence of interference,  the output random variable $\hat{B}_i$ depends on both inputs. Hence we have the Markov kernel denoted by $\mathbf{P}_{\hat{B}_I |(B_i,B_2)}$.

Practitioners can freely choose these two cases within our framework; however, as will be shown, the memoryless channel arrangement is more favored for our auditing purposes. To compare them, we need a performance metric. It is related to the following critical terms in our audit method.

Recall our previously mentioned audit method requires deriving an observation $\bar{e}$ for the random variable $\frac{1}{n}\sum_i E_i$, which is the sample mean.  It inevitably drives us to assess how ``good'' we can make such an observation and what it implies to our final derived privacy lower bound. This is done via the following bias and variance argument.

\notbf{\one The bias argument.} In Equation \eqref{equ:avg_bit_error}, we define the \textit{averaged bit error} $p^{e}$ which is the expectation for $\frac{1}{n}\sum_i E_i$. The bias term describes the distance between $p^{e}$ achieved in an audit implementation and the best achievable bit error. We can trivially derive the result $p^{e}=\sum_i p_i^e\geq p_f^e$ as a result of Theorem \ref{thm:bit_error_lower_bound}. We want $p^{e}$ as small as possible by the basic considerations mentioned in Section \ref{sec:framework}. Now a question arises: when and how $p^{e}$ equals $p_f^e$.

The following theorem result gives us a more informative result for such a question.

\begin{theorem}[Achievability of $p_f^e$, proof in Appendix \ref{app:proof_inter_reduce_mi}]\label{thm:inter_reduce_mi}
In our $(n,\frac{1}{2},\mathcal{H},\mathcal{C}_{\mathcal{M}},\mathcal{D})$ framework, with $p_i^{e}$ defined in Equation \eqref{equ:bit_error}, if $\mathcal{M}$ is $f$-DP, then we have $\forall i\in [n]$,
\begin{equation}\label{equ:inter_reduce_mi}
     1-\mathrm{h}(p_i^{e})\leq\mi(B_i;\hat{B}_i)\overset{\mathbf{1}}{\leq} \mi(B_i;\hat{B}_i|B_{-i})\leq u_f(\frac{1}{2})
\end{equation}
inequality $\mathbf{1}$ becomes equality only when $\mi(B_{-i};\hat{B}_i)=0$. 

\end{theorem}
$\mi(B_{-i};\hat{B}_i)=0$ means recovered bit $\hat{b}_i$ is independent of input bits other than the intended $b_i$ itself (i.e., there is no interference). Note that we can also derive Equation \eqref{equ:bit_error_lower_bound} based on the above result; nevertheless, the important fact told by Theorem \ref{thm:inter_reduce_mi} is that if $\mathcal{M}$ is $f$-DP, no matter how powerful the decoder $\mathcal{D}$ (where the MIA happens) is, $p_f^e$ is not achievable for $p^e$ in the presence of interference. In other words, a \textbf{non-zero bias} always exists between $p^e$ and $p_f^e$.

This violates the basic design considerations mentioned in Section \ref{sec:framework}. Based on the bias argument, arranging bits transmission in a memoryless channel is better than a channel with interference.

\notbf{\two The variance argument.} Our variance argument investigates the variance of the random variable $\frac{1}{n}\sum_i E_i$, which directly related to our confidence in our estimation. We want as low uncertainty (low variance in estimation) as possible, which applies to any other statistical estimation method.

In the case of a memoryless channel, $E_i$ is independent of $E_j$ for all $i \neq j$, which means that
\begin{equation}\label{equ:var_prod_channel}
    \mathbf{Var}\left[\frac{1}{n}\sum_i E_i\right]=\frac{\sum_i \mathbf{Var}\left[E_i\right]}{n^2}\geq \frac{p_f^e(1-p_f^e)}{n}=V_{min}
\end{equation}
Because  $p_i^e=\pr{}{E_i=1}\geq p_f^e$ implies $\mathbf{Var}\left[E_i\right]\geq p_f^e(1-p_f^e)$.

Therefore, we favor the memoryless channel arrangement as it is only possible to achieve the best variance $V_{min}$.

\vspace{0.05cm}
\notbf{Conclusion.} Relating our audit mentioned before, we need the critical term $\bar{e}$ Equation \eqref{equ:ci} to compute a final lower bound. To have a non-trivial lower bound, we need 1) $\bar{e}$ to be as close to $p_f^e$ as possible (in expectation), i.e., we want low bias;  2) we need to have lower uncertainty in the estimation, i.e., $\bar{e}$ having small variance so that we can have non-trivial confidence in our estimation. A memoryless channel arrangement is more favored based on both lenses.

\vspace{\pmgttom}
\subsection{How the Decoder Affects Audit}
The decoder $\mathcal{D}$ is where the membership inference attack happens, and we discuss how it affects the audit in the following. We can quantitatively reason about decoder $\mathcal{D}$, for instance, considering the Markov chain $B_i\rightarrow m \rightarrow \hat{B}_i$, $\mathcal{D}$ happens under transition $m \rightarrow \hat{B}_i$. Note that $\mathrm{MI}(B_i; \hat{B}_i)= \mathrm{MI}(B_i; m)-\mathrm{MI}(B_i; m|\hat{B}_i)$. Powerful $\mathcal{D}$ leads to $\mathrm{MI}(B_i; \hat{B}_i)=\mathrm{MI}(B_i;m)$ ( $\mathrm{MI}(B_i; m|\hat{B}_i)=0$), meaning that $\mathcal{D}$ may be a one-to-one mapping or sufficient statistics, allowing tight audit. Weak $\mathcal{D}$ leads to $\mathrm{MI}(B_i; \hat{B}_i)< \mathrm{MI}(B_i; m)\leq u_f(1/2)$ ($\mathrm{MI}(B_i; m|\hat{B}_i)>0$), leading to non-zero bias, which must end up with a gap between the lower and upper bound for any auditing method.

Various factors may cause $\mathcal{D}$ to be weak: the membership inference attack is just sub-optimal, or there exist random sources unknown to the adversary (just like the adversary doesn’t know which other data examples are sampled in each iteration of DP-SGD). Intuitively, $\mathcal{D}$ being weak means we lose information when processing the data. Once we know  $\mathrm{MI}(B_i; m|\hat{B}_i)$, we know how quantitatively $\mathcal{D}$ affects the audit’s tightness by Theorem \ref{thm:inter_reduce_mi}, however, computing the value for $\mathrm{MI}(B_i; m|\hat{B}_i)$ should depend on the applications.

We also emphasize that in the presence of interference discussion in the above section, $\mathrm{MI}(B_i; \hat{B}_i)< \mathrm{MI}(B_i; m)\leq u_f(1/2)$ will also be true, causing the decoder $\mathcal{D}$ to be weak. This gives another reason why we should have a memoryless channel arrangement.

% \vspace{\smgbs}
\section{Privacy Audit by One Run: (Im)possibilities}\label{sec;one_run_possible}
Now, we are prepared to answer previously raised questions about the main topic we aim to discuss: \textit{privacy audit by one run}.

\notbf{The nature of privacy audit is to estimate the randomness due to DP.} Using our bias and variance argument mentioned before, in all statistical estimation tasks, we always favor the expectation that the subject measured is close  (lower bias) to its true value and high confidence (low variance). Under the context of privacy audit, what we really want to estimate is the \textit{randomness injected by DP, which is parameterized by privacy parameters}. 

For example, for the Gaussian mechanism 
\begin{equation}\label{equ:gm}
    \mathcal{M}(X)=q(X)+\mathcal{N}(0,\sigma^2\mathbb{I}^d)
\end{equation} 
where the query function $q(X)\in\mathbb{R}^d$ has unit $\ell_2$-sensitivity, it is known that it satisfies $(\varepsilon,\delta)$-DP if $\sigma^2\geq2\log(1.25/\delta)/\varepsilon^2$. Estimating a lower bound $\varepsilon_L$ for $\varepsilon$ is equivalent to estimating an upper bound for the noise s.t.d. $\sigma$. Inevitably, we must have enough observations of \textit{independent} samples from the DP randomness itself before confidently claiming something about $\sigma$. 

In privacy audit, the observations we have is the $\{(b_t,b_g)_i:\forall i \in [n]\}$ pairs of truth and guesses (Equation \eqref{equ:guess_pairs}). The goal is to estimate the randomness of DP based on those pairs. Interference will always lead to sub-optimal results based on our bias and variance argument.

\notbf{When it is possible to audit privacy by one run.} Suppose we plan to insert $n$ canaries for audit. Privacy audit by one run is only possible if 1) the targeted private algorithm $\mathcal{M}$ itself incurs sampling from at least $n$ independent DP randomness source; 2) $n$ is large enough to have concentration behaviors.

The second requirement is because we need to have non-trivial confidence due to statistical uncertainties, and the first requirement is to have quality estimation based on our bias and variance argument. In the following, we will use a positive and negative example to give more insight into such necessary conditions for privacy audits by one run.

\notbf{Example.} In our audit framework, we aim to audit the privacy of the Gaussian mechanism defined in Equation \eqref{equ:gm}. And the query
function q is just a summation query
     
\textbf{Positive case.}  If $d\geq n$, we can audit privacy by one run of the Gaussian mechanism and maintain the more favored memoryless channel arrangement mentioned previously. The canary data example is formed as each canary data $X_i\in R^d$ takes the value of $1$ only its $i$-th coordinate and zero for the rest. The decoder $\mathcal{D}$ is also simple: for bit $b_i$, the recovered $\hat{b}_i$ is only based on the $i$-th coordinate of the output of $\mathcal{M}$. We can see that recovering $\hat{b}_i$ is free from other input bits.

\textbf{Negative case.} If $d\ll n$, inserting more than $d$ canaries will only lead to sub-optimal results according to our previous bias and variance argument, as we only have $d$ independent source for DP noise. If $d$ is too small, we can not have non-trivial confidence in our estimation. Therefore, audit-by-one-run is impossible for this case, so we have to resort to audit-by-multiple-runs to give meaningful privacy lower bound.

\begin{figure*}[!t] 
    \centering
    \subfloat[$\mu=0.2$]
    {
    \includegraphics[width=.33\linewidth]{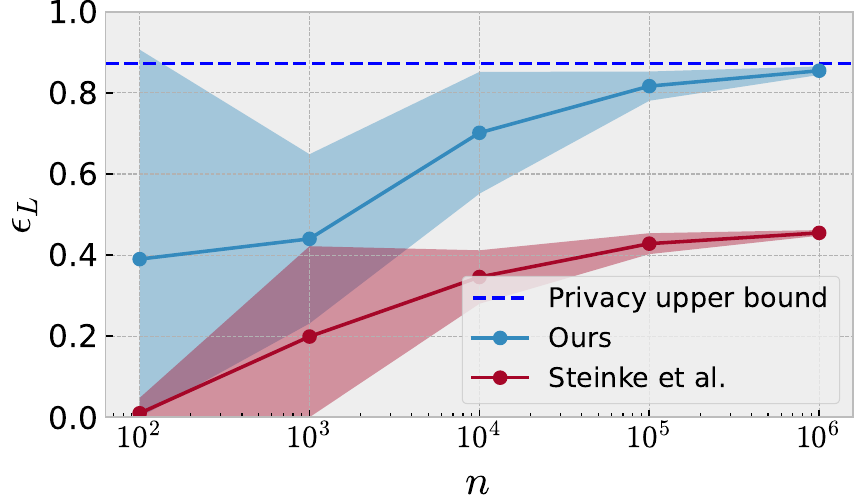}
    }
    \subfloat[$\mu=0.8$]
    {
    \includegraphics[width=.33\linewidth]{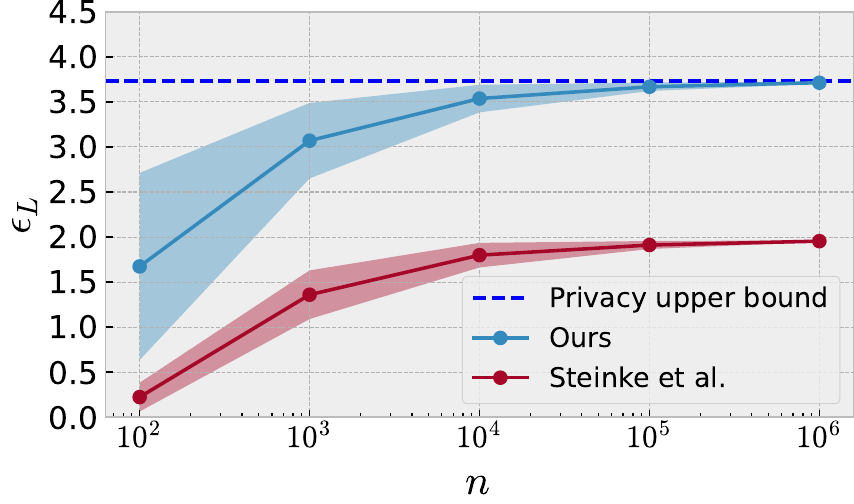}
    }
    \subfloat[$\mu=3.2$]
    {
    \includegraphics[width=.33\linewidth]{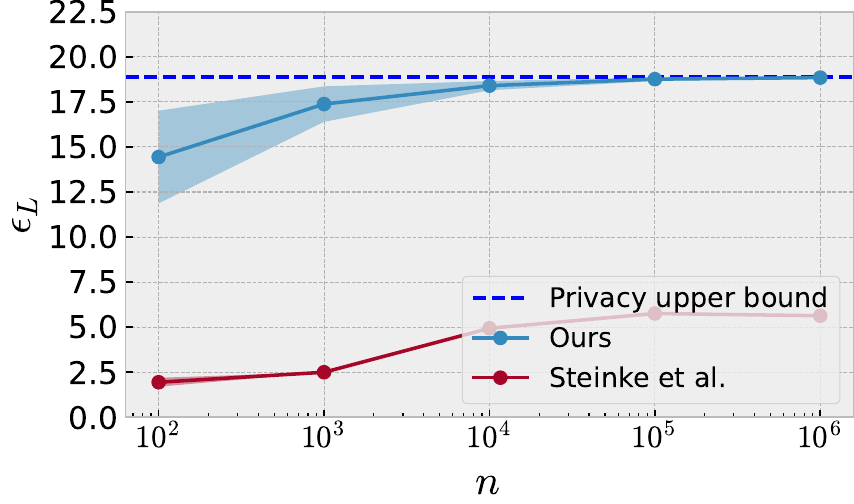}
    }

    \vspace{\fmgftoc}
    \caption{
   Audit by one run for the Gaussian mechanism satisfying $\mu$-GDP. $20$ repetition with different seeds. }
    \label{fig:gdp_audit} 
    \vspace{\fmgctom}
\end{figure*}

\begin{figure*}[!t] 
    \centering
    \subfloat[$\varepsilon=0.25$]
    {
    \includegraphics[width=.33\linewidth]{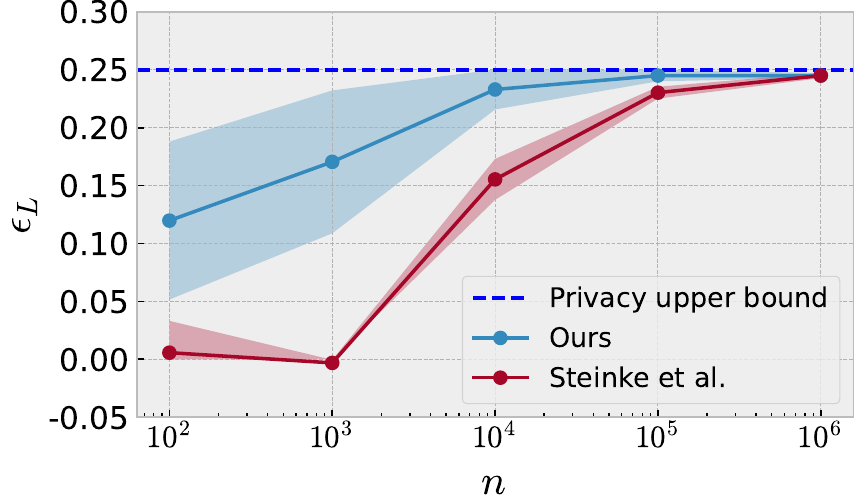}
    }
    \subfloat[$\varepsilon=1$]
    {
    \includegraphics[width=.33\linewidth]{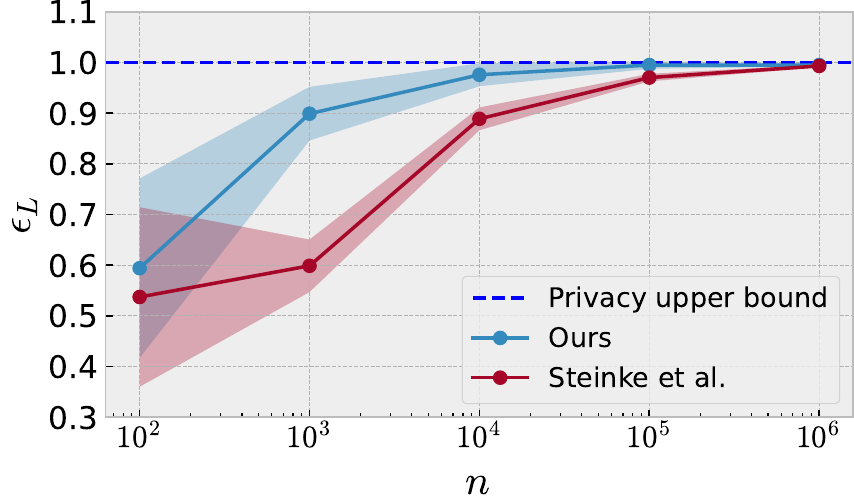}
    }
    \subfloat[$\varepsilon=4$]
    {
    \includegraphics[width=.33\linewidth]{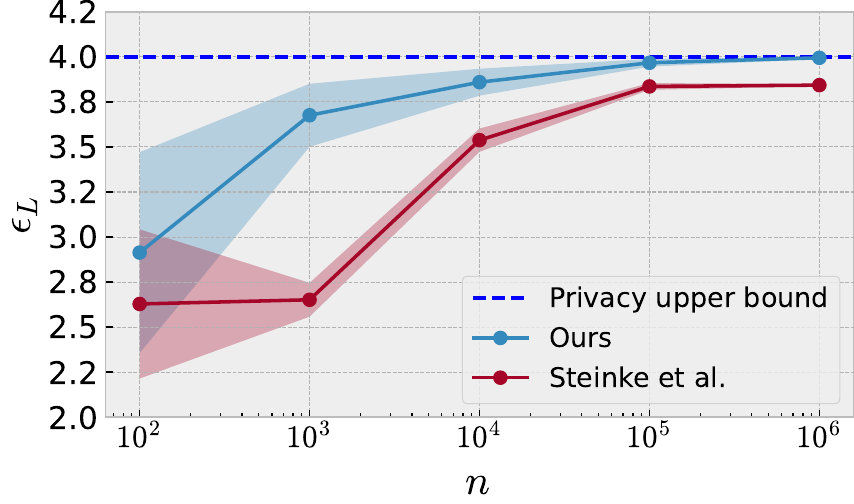}
    }

    \vspace{\fmgftoc}
    \caption{
   Audit by one run for the randomized response mechanism satisfying $(\varepsilon,10^{-5})$-DP. $20$ repetition with different seeds.
    }
    \label{fig:eps_audit} 
    \vspace{\fmgctom}
\end{figure*}

\begin{figure*}[!t] 
    \centering
    \subfloat[$\mu_l=0.2$]
    {
    \includegraphics[width=.33\linewidth]{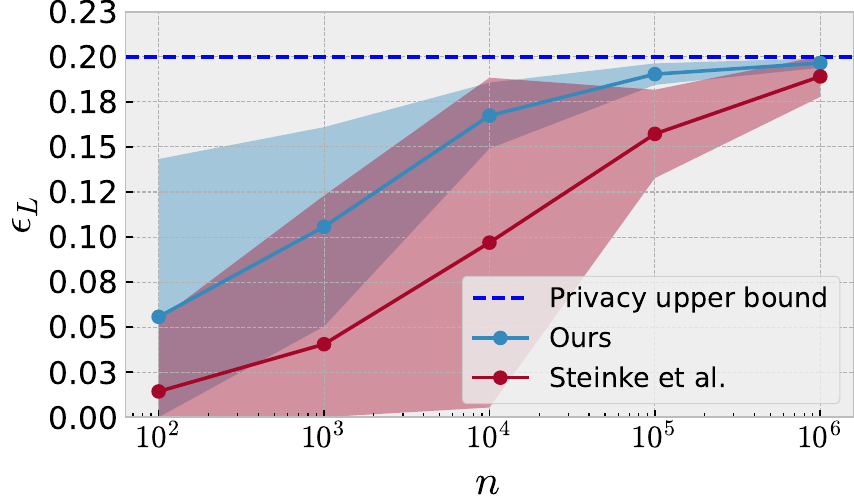}
    }
    \subfloat[$\mu_l=0.8$]
    {
    \includegraphics[width=.33\linewidth]{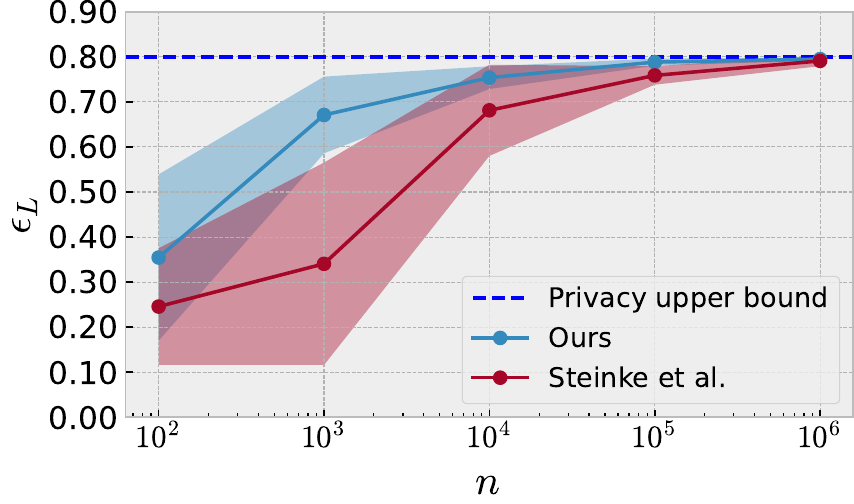}
    }
    \subfloat[$\mu_l=3.2$]
    {
    \includegraphics[width=.33\linewidth]{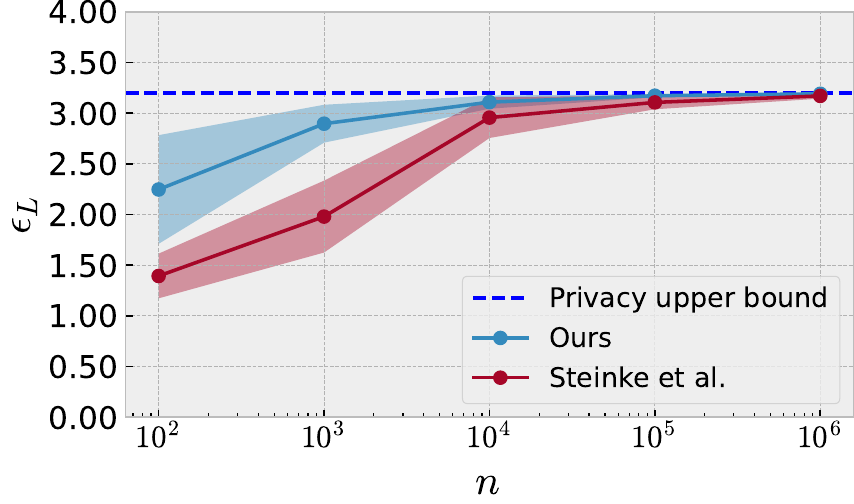}
    }

    \vspace{\fmgftoc}
    \caption{
   Audit by one run for the Laplace mechanism satisfying $T_{Lap(0,1),Lap(\mu_l,1)}$-DP. $20$ repetition with different seeds.
    }
    \label{fig:lap_audit} 
    \vspace{\fmgctom}
\end{figure*}

% \vspace{\smgbs}
\section{Experiments for Privacy Audit by One Run}
% \vspace{\pmgttom}

\subsection{Tight Audit by One Run}
In this section, we first give results showing that our audit method is indeed tight. The confidence $\gamma$ is set to be $0.95$ throughout our experiments. 

We provide three experiments in the following. 

\notbf{How the experiments fits into our $(n,\frac{1}{2},\mathcal{H},\mathcal{C}_{\mathcal{M}},\mathcal{D})$ framework.} We follow identical setups as that of \cite{steinke2023privacy}. For Gaussian mechanism, in Equation \eqref{equ:gm}, $X$ contains rows with different one-hot vectors (row number one is [1,0,0,...,0], row number two is [0,1,0,...,0], etc); $q$ sums up all vectors into one vector;  $\mathcal{D}$ looks into each coordinate of final noisy vector (after Gaussian noise added) and output each $\hat{b}_i$. For auditing the Laplace mechanism in the following, the noise is merely changed to Laplace noise compared to auditing Gaussian mechanism. For randomized response, the DP randomness is randomly flipping the bits (instead of adding Gaussian noise) and $\mathcal{D}$ output each $\hat{b}_i$ based on noisy bits.

For example, in auditing the Gaussian mechanism, $X_i$ refers to the $i$-th row of $X$, $m$ is the final vector with Gaussian noise added, and $\hat{b}_i$ is the guessed bit for $b_i$. The dataset generator $\mathcal{H}$ is also simple: depending on $b_i$, it sets $X_i$ to be a one-hot vector or a zero vector.

For asserting $\hat{b}_i$, we simply make decoder output 1 if the observed value is greater than 0.5 (the ``threshold'') for the Gaussian and Laplace mechanism. In previous work \cite{steinke2023privacy,nasr2023tight}, it is often to tune the threshold to reach the strongest audit performance. However, it suffices to set the threshold to be 0.5 in our case. For auditing the randomized response mechanism, we simply make the decoder output bit equal to the observed bit as it maximizes the posterior probability.

\notbf{1) audit the Gaussian mechanism} (Equation \eqref{equ:gm}) where we show our method obtains tight results and previous work \cite{steinke2023privacy} does not in contrast.

\textbf{Setup.} We aim to transmit $n$ bits, and we let $d=n$. We vary $n$ to take multiple values; the data example canary is according to the \textit{positive case} in Section \ref{sec;one_run_possible}. This means that we have a memoryless channel arrangement. In this experiment, the query function $q$ is just a summation query. In $f$-DP formulation, the Gaussian mechanism satisfies $\mu$-GDP. We also vary $\mu$ to see audit results in different setups.

\textbf{Results.} Figure \ref{fig:gdp_audit} shows the audit results using our method and the previous method by Steinke et al. \cite{steinke2023privacy}. And we can see that our method can achieve almost tight audit results. In contrast, the previous method cannot achieve tight results, as reported in the original work \cite{steinke2023privacy}. We believe one important reason is that \cite{steinke2023privacy} is based on $(\varepsilon,\delta)$-DP formulation, which is not tight/faithful \cite{dong2019gaussian,zhu2022optimal} for Gaussian mechanism.

By using $f$-DP formulation, we get tight results. It should also be noted that it is unclear how to transfer \cite{steinke2023privacy}'s result to handle the $f$-DP formulation.

\notbf{2) audit randomized response mechanism \cite{warner1965randomized}} in an idealized setting where \cite{steinke2023privacy} gives tight audit results, but we show that our gives tight results with $n$ less than one order of magnitude. 

\textbf{Setup.} In our experiment, we have $n$ bits to transmit, and randomized response turns the original bit into three possible outcomes: if $b_i=0$, with probability $\frac{(1-\delta)\mathrm{e}^\varepsilon}{1+\mathrm{e}^\varepsilon}$, output $0$; with probability $\frac{(1-\delta)}{1+\mathrm{e}^\varepsilon}$, output $1$; with probability $\delta$, output $2$.  If $b_i=1$, with probability $\frac{(1-\delta)\mathrm{e}^\varepsilon}{1+\mathrm{e}^\varepsilon}$, output $1$; with probability $\frac{(1-\delta)}{1+\mathrm{e}^\varepsilon}$, output $0$; with probability $\delta$, output $3$. Then $\hat{b}_i$ is guessed based on such output. It is clear that such a mechanism satisfies $(\varepsilon,\delta)$-DP. It is also clear that we have a memoryless channel arrangement.

\textbf{Results.} Figure \ref{fig:eps_audit} shows the audit results using our method and the previous method by Steinke et al. \cite{steinke2023privacy}. We see that the previous method can achieve tight audit results when $\varepsilon=0.25,1$, but a notable gap is still seen when $\varepsilon=4$. In contrast, our method achieves tight results for all setups, and we obtain tight results with $n$ being less by one order of magnitude.

\notbf{3) audit the Laplace mechanism.} The Laplace mechanism is summarized below:
$$\mathcal{M}(X)=q(X)+\mathcal{LAP}(0,c).$$ Note that $q(X)\in \mathbb{R}^d$ has bounded $l_1$-norm and  $\mathcal{LAP}(0,c)$ is a $d$-dimension Laplace noise vector (with mean equal to zero and scale parameter $c$) where each coordinate is independent of each other. 

\textbf{Setup.} We set $q(X)\in \mathbb{R}^d$ has bounded $l_1$-norm equals to 1, then the mechanism satisfies $(\varepsilon_L=1/c,0)-DP$. By re-parameterizing, the trade-off function is $T_{Lap(0,1),Lap(\mu_l,1)}$-DP where $\mu_l=1/c$.

\textbf{Results.} The results are presented in Figure \ref{fig:lap_audit}. We can see that both audit methods can achieve tight results when $n=d$ is large enough; however, to reach the same lower bound, our method is more efficient by around one order of magnitude. This conclusion is similar to what Figure \ref{fig:eps_audit} tells us.

\begin{figure}[!t] 
    \centering
    \subfloat[$n=10^{3}, \delta=10^{-5}$]
    {
    \includegraphics[width=.9\linewidth]{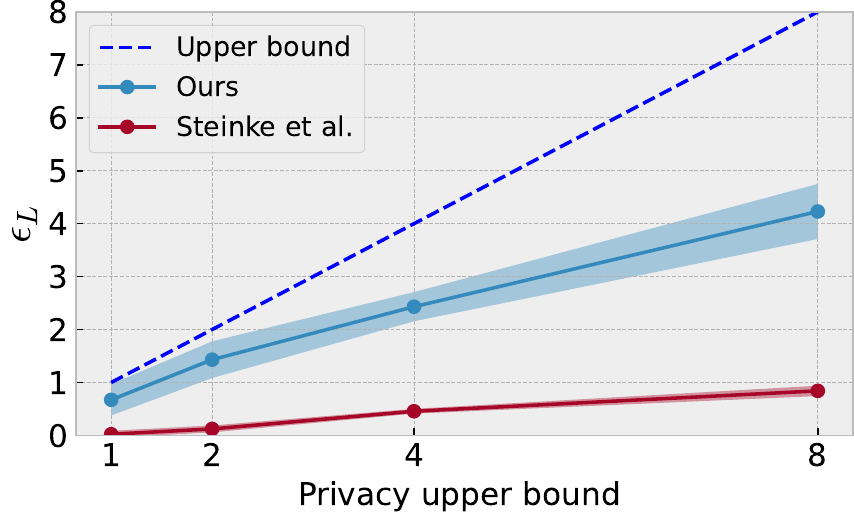}
    }\\
    \subfloat[$n=10^{4}, \delta=10^{-5}$]
    {
    \includegraphics[width=.9\linewidth]{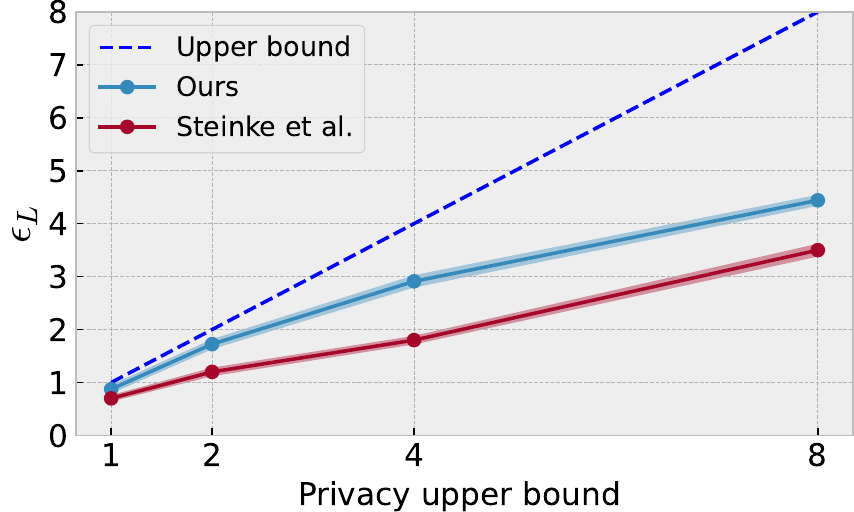}
    }\\
    
    \vspace{\fmgftoc}
    \caption{
   Privacy lower bounds for auditing DP-SGD at white-box setting. The experimental dataset is CIFAR10, the same as that in \cite{steinke2023privacy}.
    }
    \label{fig:learning_audit} 
    \vspace{\fmgctom}
\end{figure}

\vspace{\pmgttom}
\subsection{Experiments on DP-SGD}

Our main contribution in this paper is analyzing the result of membership inference, particularly for the privacy-audit-by-one-run case; therefore, we do not focus on how to launch stronger membership inference, and we leverage previous techniques for membership inference. Both our method and previous method \cite{steinke2023privacy} are based on the same membership inference result, allowing us to have fair comparisons. Experiment implementation is at a link \footnote{{ https://github.com/zihangxiang/PAABT.git}}.

In this section, the audit results for DP-SGD protocols are given. We focus on the white-box setting where the intermediate private gradient is released publicly. We also leverage the membership inference method provided by Nasr et al. called ``Dirac gradient,'' which directly inserts gradient candies where only one coordinate is 1 with others being zero. Such practice is similar to our above audit on the Gaussian mechanism, and we indeed have a memoryless channel arrangement based on a similar argument.

\notbf{Results.} Figure \ref{fig:learning_audit} presents the audit result. We can see that our method produces better lower bounds in each setting. Although the audit is not tight when performed on real-world training tasks when $n=10^{4}$, both our method and the previous method give meaningful lower bounds; however, our method has significant advantages when $n=10^{3}$. 

% \notbf{1) auditing in the white box setting} (Equation \ref{equ:gm}) where we show our method obtains tight results and previous work \cite{steinke2023privacy} does not in contrast.

\vspace{\pmgttom}
\subsection{Detecting Privacy Violation}
In this experiment, we provide a use case showing that our method catches bugs in real-world applications of differential privacy. This study is based on a pitfall in trying to refine the DP-SGD protocol. We briefly describe the root cause of such error made in \cite{stevens2022backpropagation} in the following.

The original DP-SGD protocol can be concisely summarized as 
$$p_i = \frac{1}{|\mathbf{B}|}\sum_i\operatorname{CLP}_C\left(\nabla \ell_i\right)+\textbf{Noise}$$ 
where $\nabla \ell_i$ is the per-example gradient,$|\mathbf{B}|$ is the batchsize and $\operatorname{CLP}_C(u) = u\cdot \min(1,\frac{C}{\|u\|_2})$ is a clipping operation. The final gradient $p_i$ is differential private by adding calibrated noise. However, \cite{stevens2022backpropagation}'s solution accidentally turns the above protocol into 
$$p_i = \sum_i\operatorname{CLP}_C\left(\frac{1}{|\mathbf{B}|}\nabla \ell_i\right)+\textbf{Noise}$$ which makes the added noise to be considerable underestimated. We also focus on the white-box setting for auditing such privacy bugs.

\begin{figure}[!t] 
    \centering
    \subfloat[$n=10^{3}, \delta=10^{-5}$]
    {
    \includegraphics[width=.9\linewidth]{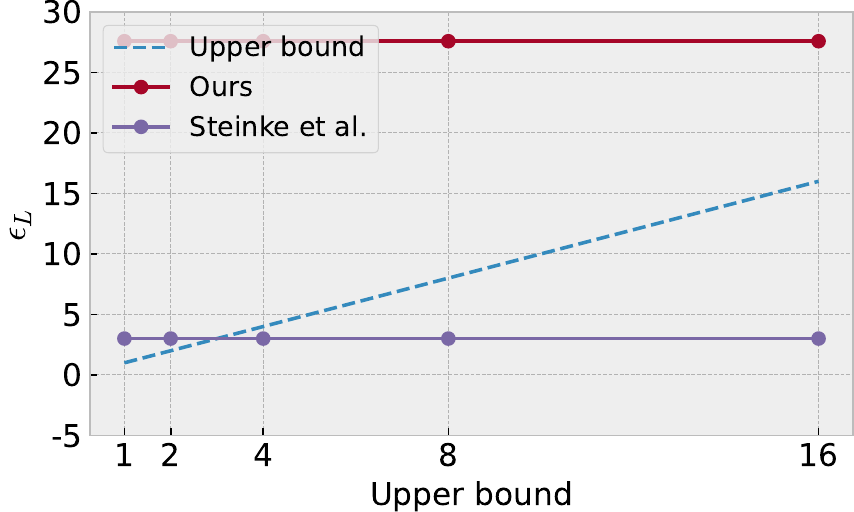}
    }\\
    \subfloat[$n=10^{4}, \delta=10^{-5}$]
    {
    \includegraphics[width=.9\linewidth]{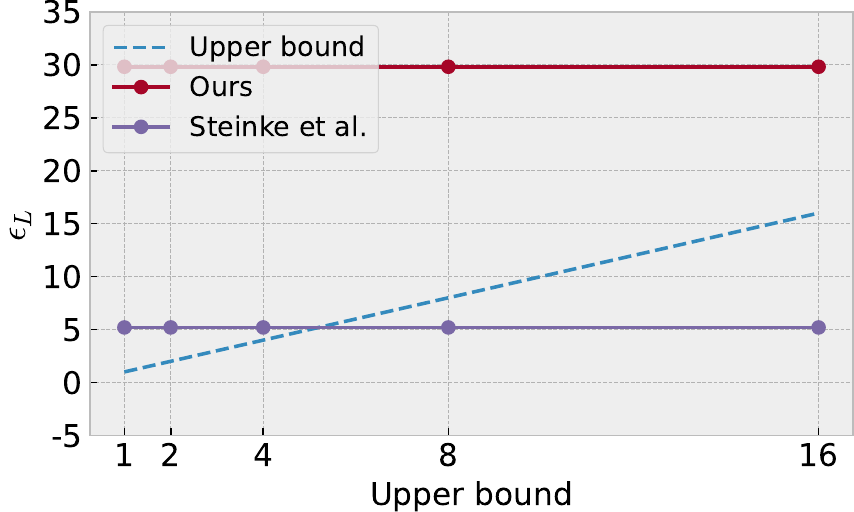}
    }
    \vspace{\fmgftoc}
    \caption{
    Auditing a flawed private algorithm implementation.
    }
    \label{fig:bug_audit} 
    \vspace{\fmgctom}
\end{figure}

\notbf{Results.} Figure \ref{fig:bug_audit} presents the audit results by both methods. Our method produces a rather strong lower bound, asserting that the algorithm is blatantly not as private as claimed under all privacy setups. In contrast, the previous method \cite{steinke2023privacy} only captures privacy violations for setups where the claimed privacy upper bounds are some small value and fails to catch the violations when they become greater.

\vspace{\smgbs}
\section{Conclusion}
\vspace{\pmgttom}

In this work, we present a unifying framework that models the privacy audit problem through the lens of information-theoretic bit transmission. Using this approach, we derive fundamental limits for privacy audits and develop an audit method that provides tight empirical privacy bounds. Our framework also facilitates a detailed exploration of auditing via a single run of the target algorithm, investigating its feasibility and inherent limitations. These contributions not only enhance the accuracy and efficiency of privacy audits but also offer deeper insights into the problem of when and how auditing by one run is possible/impossible.

Our experimental results demonstrate that the proposed methods yield tighter privacy bounds with fewer (up to one order of magnitude) observations while also identifying privacy vulnerabilities in flawed differentially private implementations. These findings have important implications for strengthening privacy guarantees in real-world applications.
% more robustly and efficiently.

\section{Ethics Considerations}
We investigate the privacy guarantees of differentially private algorithms by developing and applying audit techniques designed to verify the privacy claims of these algorithms. Our method does not instantiate any real-world privacy attacks. The techniques and findings presented in this paper are intended solely for academic purposes and to advance the field of differential privacy. 
We aim to contribute to developing more secure and trustworthy privacy-preserving technologies.

\section{Open Science}

Our paper complies with the CFP open science policy. We follow the policy to openly share the research artifacts, including the source codes to realize secure quantile summary aggregation, scripts to run the evaluations, and documents to understand the functions of code with respect to the descriptions in this paper.

% %-------------------------------------------------------------------------------
\section{Acknowledgments}
Di Wang and Zihang Xiang are supported in part by the funding BAS/1/1689-01-01, URF/1/4663-01-01,  REI/1/5232-01-01,  REI/1/5332-01-01,  and URF/1/5508-01-01  from KAUST, and funding from KAUST - Center of Excellence for Generative AI, under award number 5940. Tiaohao Wang is supported by NSF CNS-2220433.
% %-------------------------------------------------------------------------------
% \section*{Acknowledgments}
% %-------------------------------------------------------------------------------

% The USENIX latex style is old and very tired, which is why
% there's no \textbackslash{}acks command for you to use when
% acknowledging. Sorry.

% %-------------------------------------------------------------------------------
% \section*{Availability}
% %-------------------------------------------------------------------------------

% USENIX program committees give extra points to submissions that are
% backed by artifacts that are publicly available. If you made your code
% or data available, it's worth mentioning this fact in a dedicated
% section.

%-------------------------------------------------------------------------------
\bibliographystyle{plain}
\bibliography{ref}

%%
%% If your work has an appendix, this is the place to put it.
% \balance
\appendix

\section{Content for reference}

\vspace{\smgbs}
\subsection{Neyman–Pearson Lemma}\label{app:np_lemma}
\vspace{\pmgttom}
\begin{theorem}[Neyman–Pearson lemma \cite{neyman1933ix}]
Let $P$ and $Q$ be probability distributions on $\Omega$ with densities $p$ and $q$, respectively. Define $L(x)=\frac{p(x)}{q(x)}$. For hypothesis testing problem 
{\small$$\mathbf{H_0}:P,\text{\quad}\mathbf{H_1}: Q$$}For a constant $c>0$, suppose that the likelihood ratio test which rejects $\mathbf{H_0}$ when $L(x)\leq c$ has $\mathrm{FP}=a$ and $\mathrm{FN}=b$, then for any other test of $\mathbf{H_0}$ with $\mathrm{FP}\leq a$, the achievable false negative rate is at least $b$.
    
\end{theorem}

Neyman–Pearson lemma says that the most powerful test (optimal false negative rate) at fixed  false positive rate is the likelihood ratio test. 

% Applying Neyman–Pearson lemma to distinguishing $\mathcal{N}(0,1)$ from $\mathcal{N}(\mu,1)$ gives us Definition \ref{def:gdp} \cite{dong2019gaussian}.

\subsection{Hoeffding's Inequality}\label{app:hoeffding_bound}

\begin{theorem}[\cite{wiki:Hoeffding's_inequality}]\label{thm:hoeffding_bound}
    Suppose $\bar{X}= \frac{1}{n}\sum_i X_i$. where $a \leq X_i \leq b$ are independent, Then for any $t>0$,
    $$
    \pr{}{|\bar{X}-\mu| \geq t} \leq 2 \exp \left(-\frac{2 n t^2}{(b-a)^2}\right)
    $$
\end{theorem}
Note that $X_i$ does not necessarily need to be identically distributed. In our case, $a=0,b=1, \mu=p_f^e$, hence, setting 
$$2 \exp \left(-2 n v^2\right)=2\cdot(1-\gamma)$$ give us 
$v=\sqrt{\frac{1}{2n}\log\frac{1}{1-\gamma}}$.

\section{Proofs}
\subsection{Proof of Property \ref{property:bit_fdp_conditioned}}\label{app:proof_bit_fdp_conditioned}
\begin{proof}
    By fixing $B_{-i}=b_{-i}$, $\mathcal{M}$ is $f$-DP means that message distribution $M|{B_i=0, B_{-i}=b_{-i}}\fdp{f}M|_{B_i=1,B_{-i}=b_{-i}}$, because $M$ is only formed based on  $\mathcal{M}$'s output and $X_i$ is only included into exactly one run of $\mathcal{M}$.

    By post-processing property of $f$-DP, we have $$\hat{B}_i|_{B_i=0, B_{-i}=b_{-i}}\fdp{f}\hat{B}_i|_{B_i=1,B_{-i}=b_{-i}}$$ as $\hat{B}_i$ is post-processing of $m$.
\end{proof}

\subsection{Proof for Lemma \ref{lem:mix_become_harder}}\label{app:proof_mix_become_harder}
\begin{proof}
    Denote $S=\sum_{i=1}^n c_i P_i$ and $S'=\sum_{i=1}^n c_i P'_i$ as our null and alternative hypothesis. Consider an arbitrary decision rule $\mathcal{R}$ that takes a sample from $S$ or $S'$ and rejects the null. And suppose we have the false positive rate $\alpha_\mathcal{R}=\mathbb{E}_S[\mathcal{R}]$, then 
    \begin{equation}\nonumber
    \begin{aligned}
        \alpha_\mathcal{R}=&\int_{\{\mathcal{R}(x)=1\}} \sum_{i=1}^n c_i P_i(x)\mathrm{d}x= \sum_{i=1}^n c_i \int_{\{\mathcal{R}(x)=1\}} P_i(x)\mathrm{d}x
    \end{aligned}
    \end{equation}
    Let $\alpha_i= \int_{\{\mathcal{R}(x)=1\}} P_i(x)\mathrm{d}x$, which is the false positive rate achieved by rule $\mathcal{R}$ for distinguishing $P_i$ V.S. $P'$, then $ \alpha_\mathcal{R}=\sum_{i=1}^n c_i \alpha_i$.  The false negative rate of $\mathcal{R}$ distinguishing $S$ V.S. $S'$ is $\beta_\mathcal{R}=1 - \mathbb{E}_{S'}[\mathcal{R}]$. We have that
    % $\beta_\mathcal{R}=1 - \mathbb{E}_{S'}[\mathcal{R}]$
    \begin{equation}\nonumber
    \begin{aligned}
        1 - \beta_\mathcal{R}=&\mathbb{E}_{S'}[\mathcal{R}]=\int_{\{\mathcal{R}(x)=1\}} \sum_{i=1}^n c_i P'_i(x)\mathrm{d}x\\
        =&\sum_{i=1}^n c_i \int_{\{\mathcal{R}(x)=1\}} P'_i(x)\mathrm{d}x\\
        \overset{A}{\leq}&1 - \sum_{i=1}^n c_i f(\alpha_i)\\
        \overset{B}{\leq}&1 - f(\sum_{i=1}^n c_i \alpha_i)=1 - f(\alpha_\mathcal{R}).
    \end{aligned}
    \end{equation}
    This means that $ \beta_\mathcal{R}\geq f(\alpha_\mathcal{R})$, which means that  $S \overset{f-\text{DP}}{\sim} S'$ by definition.
    $A$ is because $P_i \overset{f-\text{DP}}{\sim} P'_i, \forall i=1,\cdots,n$. $B$ is because  Jensen’s inequality and trade-off function is convex.
\end{proof}

\subsection{Proof for Corollary \ref{cor:marginal_bit_recover_hard}}\label{app:proof_marginal_bit_recover_hard}

\begin{proof}
    According to Property \ref{property:bit_fdp_conditioned}, if $\mathcal{M}$ is $f$-DP, we have $\forall i\in [n], b_{-i}\in\{0,1\}^{n-1}$,
    $$\hat{B}_i|_{B_i=0,B_{-i}= b_{-i}}\fdp{f} \hat{B}_i|_{B_i=1,B_{-i}= b_{-i}},$$
    For the conditional distribution $\hat{B}_i|_{B_i=0}$ It is equal to the following distribution in convex combination form 
    $$\sum\limits_{b_{-i}\in\{0,1\}^{n-1}}\prob\left[B_{-i}=b_{-i}\right]\hat{B}_i|_{B_i=0,B_{-i}= b_{-i}}$$
    the same also applies to $\hat{B}_i|_{B_i=1}$. By Lemma \ref{lem:mix_become_harder}, we have 
    $$\hat{B}_i|_{B_i=0}\fdp{f} \hat{B}_i|_{B_i=1}.$$
\end{proof}

\subsection{Proof for Theorem \ref{thm:mi_bound_bit_trans}}\label{app:proof_mi_bound_bit_trans}
\begin{proof}
Define a hypothesis testing problem as follows.
$$\mathbf{H_0}:\text{$B_i=0$},\text{\quad}\mathbf{H_1}:\text{$B_i=1$}. $$ I.e.,  $\hat{B}_i$ the result of our hypothesis testing. For any decision rule $\mathcal{R}$, leading to false positive rate $\alpha_\mathcal{R}$, as governed by the trade-off function, we must obtain false negative rate $\beta_\mathcal{R}\geq f(\alpha_\mathcal{R})$. Because $\hat{B}_i|B_i=0\fdp{f}\hat{B}_i|B_i=1$.

Based on the above decision rule $\mathcal{R}$, we expand the mutual information quantity as follows. ($\mathrm{H}$ is the entropy function and $\mathrm{h}$ is the binary entropy function).
\begin{equation}\label{equ:mi_expand_1}
\begin{aligned}
    \mi(G;\hat{G})=&\mathrm{H}(\hat{G})-\mathrm{H}(\hat{G}|G) \\
    =&h(\prob(\hat{G}=0))-p\cdot\mathrm{h}(\prob(\hat{G}=0|_{G=1}))\\
    &-(1-p)\cdot\mathrm{h}(\prob(\hat{G}=0|_{G=0}))\\
    =&\mathrm{h}(p\beta_\mathcal{R}+(1-p)(1-\alpha_\mathcal{R}))-p\cdot\mathrm{h}(\beta_\mathcal{R})\\
    &-(1-p)\cdot\mathrm{h}(1-\alpha_\mathcal{R})\\
    \overset{\triangle}{=}& F(\alpha_\mathcal{R},\beta_\mathcal{R},p)
\end{aligned}
\end{equation}
% Because $\prob(\hat{G}=0)=\prob(\hat{G}=0, G=0)+\prob(\hat{G}=0, G=1)$
with tedious calculation, we have 
\begin{equation}\label{equ:derivative_beta}
    \frac{\partial F} {\partial \beta_\mathcal{R}}=p\log\frac{\beta_\mathcal{R}-\beta_\mathcal{R}t}{t-\beta_\mathcal{R}t}
\end{equation}
where $t=p\beta_\mathcal{R}+(1-p)(1-\alpha_\mathcal{R})$, we want to show that $\frac{\partial F} {\partial \beta_\mathcal{R}}\leq 0, \forall \alpha_\mathcal{R}, \beta_\mathcal{R}$ governed by the trade-off function. We only need to show that $\frac{\beta_\mathcal{R}-\beta_\mathcal{R}t}{t-\beta_\mathcal{R}t}\leq 1$. Expand this inequality, all boils down to check if $(1-p)(1-\alpha_\mathcal{R}-\beta_\mathcal{R})\geq 0$, which is true because $\alpha_\mathcal{R}+\beta_\mathcal{R}\leq 1$ as governed by the trade-off function.

Hence, we have 

\begin{equation}\nonumber
\begin{aligned}
    F(\alpha_\mathcal{R},\beta_\mathcal{R},p)\leq& F(\alpha_\mathcal{R},f(\alpha_\mathcal{R}),p)\\
    \leq & \max_{x\in[0,1]} F(x,f(x),p)\\
    \overset{def}{=}&\max_{x\in[0,1]} F_f(x,p)
\end{aligned}
\end{equation}
which is our result in Theorem \ref{thm:mi_bound_bit_trans}.
\end{proof}

\subsection{Proof for Theorem \ref{thm:bit_error_lower_bound}}\label{app:proof_bit_error_lower_bound}
\begin{proof}
We follow the setups in proof for Theorem \ref{thm:mi_bound_bit_trans} in Appendix \ref{app:proof_mi_bound_bit_trans}. 

% But we need to expand the mutual information the other way compared to Equation \eqref{equ:mi_expand_1}. when $p=\frac{1}{2}$.
\begin{equation}\label{equ:mi_expand_2}
\begin{aligned}
    \mi(B_i;\hat{B}_i)=&\mathrm{H}(B_i)-\mathrm{H}(B_i|\hat{B}_i) \\
    =&h(\frac{1}{2})-\mathrm{H}(E_i|\hat{B}_i) \\
    =&1 -\mathrm{H}(E_i|\hat{B}_i) \\
    \geq&1-\mathrm{H}(E_i)\\
    =&1-\mathrm{h}(p_i^e)
\end{aligned}
\end{equation}
As conditioning reduces entropy. Combining the fact that $\mi(B_i;\hat{B}_i)\leq u_f(\frac{1}{2})$ due to  Theorem \ref{thm:mi_bound_bit_trans}, we get the result we want in Theorem \ref{thm:bit_error_lower_bound}.

\end{proof}

\subsection{Proof for Theorem \ref{thm:bit_error_dominance_independent}}\label{app:proof_bit_error_dominance_independent}

\begin{proof}
    For independent Bernoulli random variables $X_1\sim\mathbf{Bernoulli}(a)$ and $Y_1\sim\mathbf{Bernoulli}(b)$, if  $a\geq b$, we have 
    \begin{equation}\label{equ:sto_don}
        \pr{}{X_1\geq t}\geq   \pr{}{Y_1\geq t}, \forall t\in \mathbb{R} 
    \end{equation}
    for random variable $X_i, Y_i$ satisfying Equation \eqref{equ:sto_don}, we call $X_i$ stochastically dominates $Y_i$.

    For all $t\in \mathbb{R}$, if $X_i$ stochastically dominates $Y_i$ for $i=1,2$, we have
    \begin{equation}\label{equ:sto_don_two_var}
    \begin{aligned}
        \pr{}{X_1+X_2\geq t}=& \mathbb{E}_{X_1}\left[\pr{X_2}{X_2\geq t-X_1}\right]\\
        \geq& \mathbb{E}_{X_1}\left[\pr{Y_2}{Y_2\geq t-X_1}\right]\\
        =& \mathbb{E}_{Y_2}\left[\pr{X_1}{X_1\geq t-Y_2}\right]\\
        \geq& \mathbb{E}_{Y_2}\left[\pr{Y_1}{Y_1\geq t-Y_2}\right]\\
        =&\pr{}{Y_1+Y_2\geq t}
    \end{aligned}
    \end{equation}
    i.e., $X_1+X_2$ also  stochastically dominates $Y_1+Y_2$, by induction, we have  $\pr{}{\sum_i X_i\geq t}\geq   \pr{}{\sum Y_i\geq t}, \forall t\in \mathbb{R} $ if $X_i$ stochastically dominates $Y_i$ for $i\in[n]$ and all $X_i, Y_i$ are mutually independent.
    As $p_i^e>p_f^e, \forall i\in[n]$, setting $X_i=E_i$ and $Y_i=S_i$ $\forall i\in[n]$ gives us the result in Theorem \ref{thm:bit_error_dominance_independent}.
    
\end{proof}

\subsection{Proof for Theorem \ref{thm:inter_reduce_mi}}\label{app:proof_inter_reduce_mi}
\begin{proof}
    The first inequality is already proven in Theorem \ref{thm:bit_error_lower_bound}, the last inequality is because $\mathrm{MI}(B_i;\hat{B}_i|B_{-i})=\mathbb{E}_{B_{-i}}\left[\mathrm{MI}(B_i;\hat{B}_i)\right]\leq u_f(\frac{1}{2})$, as $\mathrm{MI}(B_i;\hat{B}_i)\leq u_f(\frac{1}{2})$ by theorem \ref{thm:mi_bound_bit_trans}.

    For random variable $X,Y,Z$, using the chain rule of mutual information, we have 
    \begin{equation}
    \begin{aligned}
        \mathrm{MI}(X;Y,Z)=&\mi(X;Y|Z)+\mi(X;Z)\\
        =& \mathrm{MI}(X;Z,Y)\\
        =&\mi(X;Z|Y)+\mi(X;Y)
    \end{aligned}        
    \end{equation}
    if we have $X$ is independent $Z$ (which means $\mi(X;Z)=0$), the above equation give us $\mi(X;Y|Z)\geq \mi(X;Y)$.
    Setting $B_i=X, \hat{B}_i=Y, B_{-i}=Z$ give us the result for the model inequality of Equation \eqref{equ:inter_reduce_mi}. 
\end{proof}

%%%%%%%%%%%%%%%%%%%%%%%%%%%%%%%%%%%%%%%%%%%%%%%%%%%%%%%%%%%%%%%%%%%%%%%%%%%%%%%%

%%%%%%%%%%%%%%%%%%%%%%%%%%%%%%%%%%%%%%%%%%%%%%%%%%%%%%%%%%%%%%%%%%%%%%%%%%%%%%%%

%%  LocalWords:  endnotes includegraphics fread ptr nobj noindent
%%  LocalWords:  pdflatex acks

%%%%%%%%%%%%%%%%%%%%%%%%%%%%%%%%%%%%%%%%%%%%%%%%%%%%%%%%%%%%%%%%%%%%%%%%%%%%%%%%
\end{document}